\documentclass[journal]{IEEEtran}

\usepackage{cite} 
\usepackage{CJKutf8}
\usepackage{algorithm}
\usepackage{algorithmic}
\usepackage{amsmath}
\usepackage{graphicx}
\usepackage{amssymb}
\usepackage{stfloats}
\usepackage{bm}
\usepackage{booktabs}
\usepackage{caption}
\usepackage{titlesec}

\usepackage{epsfig,subfigure,upgreek,caption}
\usepackage{hyperref}
\usepackage{xcolor}  
\hypersetup{
	colorlinks=true,
	linkcolor=blue,         
	citecolor=blue,         
	urlcolor=blue          
}

\ifCLASSINFOpdf

\else

\fi

\begin{document}
\title{AFDM-FTN: A Spectrally Efficient Waveform for High-Mobility Communications}
\author{Xianle Dai, Qu Luo,~\IEEEmembership{Member,~IEEE,} Jianguo Li, Fabien Héliot,~\IEEEmembership{Senior Member,~IEEE,} Shuangyang Li,~\IEEEmembership{Member,~IEEE,} Lixia Xiao,~\IEEEmembership{Member,~IEEE,}  and Pei Xiao,~\IEEEmembership{Senior Member,~IEEE}
\thanks{Xianle Dai is with the School of Information and Electronics, Beijing Institute of Technology, Beijing 100081, China (e-mail: daixl212@bit.edu.cn).}
\thanks{Qu Luo, Fabien Héliot and Pei Xiao are with the 5G/6G Innovation Centre, University of Surrey, GU2 7XH Guildford, U.K. (e-mail: \{q.u.luo, f.heliot, p.xiao\}@surrey.ac.uk).}
\thanks{Jianguo Li is with the School of Cyberspace Science and Technology, Beijing Institute of Technology, Beijing 100081, China (e-mail: jianguoli@bit.edu.cn).}
\thanks{Shuangyang Li is with the Electrical Engineering and
Computer Science, Technical University of Berlin, Berlin, 10587, Germany
(e-mail: shuangyang.li@tu-berlin.de).}
\thanks{Lixia Xiao is with Research Center of 6G Mobile Communications, School of Cyber
Science and Engineering, Huazhong University of Science and Technology,
Wuhan 430074, China   (e-mail:lixiaxiao@hust.edu.cn.)}
}

\maketitle
\begin{abstract}
This paper proposes an affine frequency division multiplexing (AFDM)-aided faster-than-Nyquist (FTN) waveform, termed AFDM-FTN, to enhance spectral efficiency (SE) in high-mobility communication scenarios. We first derive the AFDM-FTN input–output relationship and analyze the FTN-induced interference pattern in AFDM-FTN. To address the  channel estimation challenges, a low-complexity channel estimator based on the basis expansion model (BEM) is developed. By exploiting the intrinsic characteristics of the AFDM channel matrix and the FTN coefficient matrix, a multi-layer message passing (MLMP) algorithm is proposed that leverages the sparsity of the time-domain (TD) channel and the FTN coefficient matrix, where belief messages are iteratively propagated across the TD channel, FTN, and transform layers. Building upon the BEM-assisted channel estimation and MLMP, a low-complexity joint channel estimation and data detection scheme (BEM-MLMP-JCED) is further developed to iteratively refine channel estimation with the aid of transmitted data. Finally, the channel estimation lower bound, the mean square error (MSE) performance of the BEM-MLMP-JCED, and the computational complexity are analyzed. Simulation results demonstrate that the proposed AFDM-FTN system with BEM-MLMP-JCED achieves comparable BER to conventional AFDM while providing enhanced SE and reduced complexity compared to benchmark receivers.

\end{abstract}


\IEEEpeerreviewmaketitle

\vspace{-0.5em}
\section{Introduction}

The next-generation wireless communication systems demand enhanced user-experienced data rates, ultra-high reliability, massive connection density, and improved spectral efficiency (SE) \cite{sui2025multi}, \cite{rouAffineFrequencyDivision2026}. These requirements become particularly challenging in high-mobility scenarios, such as high-speed railways, connected vehicles, high-altitude platform stations (HAPS), and low-Earth-orbit (LEO) satellites, where rapidly time-varying channels   prevail \cite{wang2025afdm}. Under such highly dynamic conditions, the widely adopted orthogonal frequency-division multiplexing (OFDM) waveform suffers from inter-carrier interference (ICI) due to loss of orthogonality, leading to significant performance degradation \cite{yin2025ofdm}. 
Against this background, a new generation of advanced waveforms, such as orthogonal time frequency space (OTFS) \cite{xiao2021overview,deng2025unifying}, orthogonal delay division multiplexing (ODDM) \cite{lin2022orthogonal}, orthogonal chirp division multiplexing (OCDM) \cite{ouyang2016orthogonal}, affine frequency division multiplexing (AFDM) \cite{bemaniAffineFrequencyDivision2023}, and many others, have emerged as a promising pathway toward 6G and beyond.

As a representative multicarrier waveform for high-mobility channels,  AFDM  employs mutually orthogonal chirp subcarriers to modulate information symbols via the inverse discrete affine Fourier transform (IDAFT).  By appropriately tuning the chirp rate according to the maximum delay and Doppler spreads of the channel, AFDM can effectively separate multipath components in the affine Fourier transform (AFT) domain, thereby fully exploiting channel diversity in doubly selective channels \cite{zhu2023design}. As a result, AFDM achieves performance comparable to OTFS modulation, while significantly outperforming conventional OFDM in high-mobility scenarios. It is also worth noting that IDAFT/DAFT can be regarded as a generalization of the widely adopted inverse discrete Fourier transform (IDFT)/discrete Fourier transform (DFT). Consequently, AFDM maintains backward compatibility with OFDM while providing enhanced Doppler resilience.

\vspace{-0.5em}
\subsection{Related Works}
Due to the appealing advantages, AFDM has been extensively investigated, ranging from     channel estimation \cite{11012133}, \cite{11121670}, \cite{yinDiagonallyReconstructedChannel2024},  multiple access,   advanced receiver design \cite{luoAFDMSCMAPromisingWaveform2024}, \cite{11150613}, \cite{11185309} and integrated sensing and communications (ISAC) \cite{ranasingheJointChannelData2025},\cite{10439996}, \cite{10858612}.      
\textcolor{blue}{For example, an embedded pilot-assisted   channel estimation scheme  was initially proposed in \cite{bemaniAffineFrequencyDivision2023}. }
Later  on,   a low-complexity diagonally reconstructed channel estimation method for multi-antenna AFDM systems was proposed based on the embedded pilot structure. More recently, the authors in \cite{wuPerformanceAnalysisBEMBased2025} proposed
a generalized complex exponential basis expansion model
(GCE-BEM)-assisted channel estimation scheme to address
the challenges of the fractional frequency dispersion and
computational complexity of channel estimation.  
In addition, several detection algorithms have been proposed to enhance the performance of AFDM. These include a maximum ratio combining (MRC) scheme\cite{bemaniAffineFrequencyDivision2023}, \cite{pengResourceAllocationUplink2023}, an iterative minimum mean square error (MMSE) receiver that balances computational complexity and detection performance\cite{liChirpParameterSelection2025}, as well as a joint sparse graph design that fully exploits the sparse structure of the DAFT-domain channel and the low-density parity-check code (LDPC)    structure\cite{11150613}. 


In parallel with the requirement for strong Doppler resilience, 6G wireless systems also demand high  SE  to support ever-increasing data rates. 
To this end, AFDM has been integrated with sparse code multiple access (AFDM-SCMA) to enable multiuser transmission with enhanced SE \cite{luoAFDMSCMAPromisingWaveform2024}. 
In addition, AFDM has been combined with index modulation (IM) and generalized spatial modulation to further improve spectral efficiency\cite{11185315}. Several IM strategies, including subcarrier activation and joint symbol-index mapping \cite{zhuDesignPerformanceAnalysis2024}, have been investigated within the AFDM framework. More recently, inspired by spectrally efficient frequency division multiplexing, which improves spectral utilization by employing non-orthogonal, overlapped subcarriers in the frequency domain, a non-orthogonal AFDM waveform has been proposed. Specifically, by introducing a bandwidth compression factor into the AFDM modulation process, the proposed non-orthogonal waveform achieves higher SE while retaining AFDM’s robustness in high-mobility environments\cite{yi2025non}.

Beyond these approaches, faster-than-Nyquist (FTN) has emerged as a promising paradigm for high-SE waveform design \cite{liCodeBasedChannelShortening2020}, \cite{liTimeDomainVsFrequencyDomain2020}. By intentionally violating the Nyquist criterion and allowing controlled inter-symbol interference (ISI), FTN signaling significantly improves spectral utilization and has been widely recognized as a key enabling technique for 6G systems with stringent SE requirements. For example, the authors in  \cite{yuanIterativeReceiverDesign2020} integrated  FTN  signaling with  SCMA  and leveraged message passing algorithms to support massive connectivity with high spectral efficiency. Tong et al. combined FTN signaling with adaptive modulation schemes, including OFDM   and    OTFS, to enhance SE over high-mobility channels       \cite{tongAdaptiveFTNSignaling2025}. More recently, the authors in \cite{hongPrecodedFasterThanNyquistSignaling2025}    developed a precoding-based power allocation and a receiver framework for FTN-enabled OTFS systems for   doubly selective channels.

\vspace{-0.5em}
\subsection{Motivations and Contributions}

\textcolor{blue}{Against the aforementioned background, AFDM-FTN is particularly attractive for high-mobility channels as it combines the Doppler resilience of AFDM with the spectral-efficiency gain of FTN signaling. Meanwhile, the resulting coupling among channel time variation, FTN-ISI, and data detection calls for a structured low-complexity receiver. The motivations of this work are twofold:}
1) As  FTN  signaling can enhance  SE, a fundamental investigation into the amalgamation of FTN and  AFDM, referred to as AFDM-FTN, is essential for supporting high-mobility scenarios; \textcolor{blue}{ 2) Although FTN signaling improves SE by deliberately introducing controlled ISI, such ISI becomes more challenging in  doubly selective channels because it is coupled with both delay spread and Doppler-induced time variation. This coupling makes AFDM-FTN a jointly structured waveform rather than a direct superposition of AFDM and FTN.} Therefore, the design of advanced receiver architectures that jointly exploit the intrinsic characteristics of both AFDM and FTN is of vital importance to fully unlock  the advantages of AFDM-FTN.

The main contributions of this work are summarized as follows:
\begin{itemize}
\item  A novel AFDM-FTN system is proposed to support spectrally efficient transmission in high-mobility scenarios. By compressing the duration of the DAFT-domain signal and designing practical pulse-shaping schemes, FTN signaling is effectively incorporated into the AFDM framework. Built upon the proposed AFDM-FTN system, we derive the corresponding input-output relationship and provide an in-depth analysis of how FTN signaling affects the  resultant effective channel matrix.

\item  We develop a BEM-assisted channel estimation framework that exploits the intrinsic characteristics of the AFDM channel and FTN matrices to reformulate the nonlinear coupling between path coefficients and Doppler shifts into a compact linear model. We show that the proposed design avoids excessive pilot overhead and ill-conditioned issues encountered in conventional channel estimation methods, while significantly reducing the overall channel estimation complexity. 

\item We propose an advanced multi-layer message passing (MLMP) detector that propagates belief messages across the AFDM transform layer, the FTN layer, and the time-domain (TD) channel layer. By doing so, the inherent sparsity of both the TD channel matrix and the FTN coefficient matrix can be fully exploited to enable low-complexity detection. The detailed message propagation rules among the three layers are explicitly formulated. Building upon the proposed MLMP detector, we further develop a joint iterative data-aided channel estimation and data detection (JCED) scheme, referred to as BEM-MLMP-JCED, which achieves superior bit error rate (BER) performance with reduced computational complexity.

\item Finally, we derive a lower bound on the channel estimation error to rigorously characterize the achievable estimation accuracy. In addition, the mean squared error (MSE) performance of the proposed MLMP algorithm is analyzed using state evolution, providing theoretical insights into its convergence behavior. Moreover, a detailed computational complexity analysis of the proposed BEM-MLMP-JCED algorithm is conducted, demonstrating the superiority of the proposed approach.
\end{itemize}
\vspace{-0.5em}
\subsection{Organization}
The remainder of this paper is organized as follows: Section \ref{sec:2} presents the AFDM-FTN system model. 
Section \ref{sec:4} analyzes the input-output relationship of AFDM-FTN for channel estimation and details  the proposed joint BEM channel estimation and MLMP data detection algorithm. Section \ref{sec:5} provides the mathematical analysis, including the lower bound for channel estimation error, the MSE performance  based on state evolution, and the complexity of the entire system. Section \ref{sec:6} reports and discusses the simulation results. Section \ref{sec:7} concludes this paper.
\vspace{-0.5em}
\subsection{Notation}
Scalars, vectors, and matrices are represented in italics $a$, bold lowercase upright $\mathbf{a}$, and bold uppercase upright $\mathbf{A}$, respectively. $\mathbb{C}^{k\times n}$ denotes the $(k \times n)$-dimensional complex space. $\|\cdot \| $ denotes the Euclidean norm of a vector, $\Vert \cdot\Vert _F $ denotes the Frobenius norm of a matrix, $\mathbb{E}[\cdot]$ denotes the expectation of a vector,  $(\cdot)^*$ denotes the conjugate of an element, and $| \cdot |$ denotes the modulus of a complex number. ${\rm{diag}}(\mathbf{a})$ represents a diagonal matrix with vector $\mathbf{a}$ and ${\rm{diag}}(\mathbf{A})$ represents the operator that extracts the diagonal elements of matrix $\mathbf{A}$. $\otimes$ denotes the Kronecker product.  $\odot$ denotes the Hadamard product. $\mathbf{a}\setminus a$ denotes all variables in $\mathbf{a}$ except $a$. ${\rm{Tr}}(\mathbf{A})$ denotes the trace of $\mathbf{A}$. $(\cdot)_N$ is the modulo $N$ operation. $(\cdot)^T$ and $(\cdot)^H$ denote the transpose and the Hermitian transpose operation, respectively.

\begin{figure*}[!t] \centerline{\includegraphics[width=0.8\textwidth]{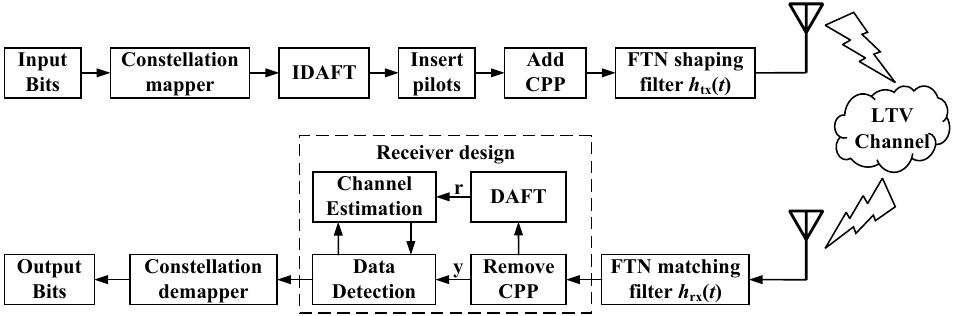}}
	\caption{Transceiver block diagram of the proposed AFDM-FTN system.}
	\label{fig1}
    \vspace{-0.5cm}
\end{figure*}
\vspace{-0.5em}
\section{System Model}
\label{sec:2} 
In this section, 
we present the system model of the proposed AFDM-FTN, as shown in Fig. \ref{fig1}.
\vspace{-0.5em}
\subsection{AFDM-FTN Modulation}
Let $ \mathbf{x}=[x[0],x[1],\cdots,x[N-1]]^T\in\mathbb{M}^{N\times1}$ be the DAFT-domain transmission symbol, where $\mathbb{M}$ represents the modulation alphabet. 
After applying IDAFT, the resulting TD samples with sampling period $T_s$, denoted as  ${\mathbf{\bar x}}=[\bar{x}[0],\bar{x}[1],\cdots,\bar{x}[N-1]]^T$,  are given by
 \begin{equation}
	\scalebox{0.9}{$%
		\begin{aligned}
			\bar{x}[n] = \sum_{m=0}^{N-1}x[m]\phi_n[m],
		\end{aligned}
		$}
	\label{eq:01}   
\end{equation}
where $\phi_n[m] = \frac{1}{\sqrt{N}}e^{j2\pi(c_1n^2 + mn/N +c_2m^2)}$, and $c_1$ and $c_2$ are the AFDM modulation parameters. Note that \eqref{eq:01} can be written in matrix form as $\bar{\mathbf{x}} = \mathbf{A}^H\mathbf{x}$,
where $\mathbf{A} = \mathbf{\Lambda}_{c_2}\mathbf{F}\mathbf{\Lambda}_{c_1}$ is the DAFT matrix, $\mathbf{\Lambda}_{c} = {\rm{diag}} \{e^{-j2\pi cn^2},n = 0,1,\cdots,N-1\}$, and $\mathbf{F}$ is the DFT matrix with its entry at the $m$th row and $n$th column given by $e^{-j2\pi mn/N}/ \sqrt{N}$.

In order to combat multipath propagation, a chirp-periodic prefix (CPP) is used in the AFDM system, such that 
\begin{equation}
	\scalebox{0.9}{$%
		\begin{aligned}
		\bar{x}[n] = \bar{x}[N+n]e^{-j2\pi c_1(N^2+2Nn)},n = -L_p,\cdots,-1,
		\end{aligned}
		$}
	\label{eq:041}   
\end{equation}
where $L_p$ is the CPP length.
%
After applying IDAFT, the AFDM symbols are passed through a shaping filter $h_{\rm{tx}}(t)$ with a period of $T_0 =  T_s/\alpha$,  i.e.,
 \begin{equation}
	\scalebox{0.9}{$%
		\begin{aligned}
			\textcolor{blue}{\bar s(t) = \sum_{k=-L_p}^{N-1}\bar{x}[k]h_{\rm{tx}}(t-k\alpha T_0),}
		\end{aligned}
        $}
	\label{eq:03}
\end{equation}
where $\alpha\in (0,1]$ is the FTN compression factor.
\subsection{Receiver model}
In this paper, we consider a doubly selective channel with the channel response at time $t $ and delay $\ell$ given by
\begin{equation}
	\scalebox{0.9}{$%
		\begin{aligned}
			h(t,\ell) = \sum_{i=0}^{L-1}h_ie^{-j2\pi f_it}\delta(\ell-l_i),
		\end{aligned}
		$}
	\label{eq:04}   
\end{equation}
where $h_i$ denotes the channel coefficient,  $f_i$ is the Doppler shift in analog frequency, the non-negative integer $ l_i$ is the $i$th delay normalized by $T_s$, $\delta(\cdot)$ is the Dirac delta function, and  $L$ is the number of paths. 
Consequently, the received signal can be expressed as
 \begin{equation}
	\scalebox{0.9}{$%
		\begin{aligned}
			y(t) = \sum_{i=0}^{L-1}\sum_{k=-\infty}^{+\infty}h_ie^{-j2\pi f_it}\bar s(t-l_iT_s)h_{\rm{rx}}(t-k\alpha T_0)+w(t),
		\end{aligned}
		$}
	\label{eq:05}   
\end{equation}
where $h_{\rm{rx}}(t)$ is a matching filter with a period of $T_s$, and $w(t)$ is a complex Gaussian white noise with zero mean and variance  $\sigma^2$. 
The shaping pulse $h_{\rm{tx}}(t)$ can be expanded over a set of
orthogonal shifts of $h_{\rm{rx}}(t)$, 
i.e., $h_{\rm{tx}}(t) = \sum{g[k]}h_{\rm{rx}}(t-k\alpha T_0)$, where $ g[k] = \int_{-\infty}^{\infty}h_{\rm{tx}}(t)h_{\rm{rx}}(t-k\alpha T_0){\rm{d}}t$ \cite{prljaReducedComplexityReceiversStrongly2012}.

Let us define  $\hat{f}_i \triangleq f_i/\Delta f \in [-\hat{f}_{\max}, \hat{f}_{\max}]$ as the Doppler shift normalized by the subcarrier spacing $\Delta f$, and $\hat{f}_{\max}$ denotes the maximum normalized Doppler. It can be further decomposed as $ \hat{f}_i= \varphi_i + \psi_i$, where $\varphi_i\in[-\varphi_{\rm{max}}, \varphi_{\rm{max}}]$ denotes the integer Doppler part and $\psi_i\in(-\frac{1}{2}, \frac{1}{2}]$ denotes the fractional Doppler part. After sampling at a rate of  $1/T_s$ and removing the CPP, the TD received signal can be expressed as
 \begin{equation}
	\scalebox{0.9}{$%
		\begin{aligned}
			y[n] = \sum_{i=0}^{L-1}h_ie^{-j\frac{2\pi}{N} \hat{f}_i n}s[n]+w[n],
		\end{aligned}
		$}
	\label{eq:06}   
\end{equation}
\textcolor{blue}{where $s[n] = \sum_{k=-\infty}^{\infty}\bar x[k]g[n-k-l_i]$ denotes the sampled FTN signal observed through the delay of the $i$th propagation path.} For notational simplicity, we will henceforth write $s[n]$ as $s_n$, and adopt the same convention for other variables. Note that \eqref{eq:06} can also be  expressed in matrix form as
 \begin{equation}
	\scalebox{0.9}{$%
		\begin{aligned}
			\mathbf{y}  = \mathbf{HGA}^H\mathbf{x} + \mathbf{w},
		\end{aligned}
		$}
	\label{eq:061}   
\end{equation}
\textcolor{blue}{where $\mathbf{H} =\sum_{i=0}^{L-1}h_i{\bm{\Delta}}_{\hat{f}_i}{\mathbf{\Pi}}^{l_i}$ denotes the TD effective channel matrix, $\mathbf{w}$ is the noise vector, ${\bm{\Delta}}_{\hat{f}_i} = {\rm{diag}}\{e^{-j\frac{2\pi}{N} \hat{f}_i n},n = 0,1,\cdots,N-1\}$ models the Doppler, $\mathbf{\Pi}$ is a cyclic shift matrix \cite{bemaniAffineFrequencyDivision2023}.
Since $g[k]$ is time localized in practical pulse shaping filters, only the dominant coefficients within the effective FTN-ISI memory order $N_I$ are retained to construct $\mathbf{G}$. Accordingly, the finite dimensional FTN coefficient matrix is given by}
\begin{equation}
\scalebox{0.9}{$%
 			\mathbf{G}=
            {\setlength{\arraycolsep}{2.0pt}
			\begin{pmatrix}
				g[0]  & g[1]  & \cdots & g[-N_I] & g[-N_I+1] & \cdots & g[-1] \\[2pt]
				g[-1]  & g[0]  & \cdots & 0 & g[-N_I] & \cdots & g[-2] \\[2pt]
				g[-2] &g[-1]  & \cdots & 0 & 0  & \cdots& g[-3] \\[2pt]
				\vdots  &\vdots  & \ddots & \vdots & \vdots & \ddots & \vdots \\[2pt]
				g[1] &g[2]  & \cdots & 0 & 0  & \cdots & g[0]
			\end{pmatrix}}
			$}
	\label{eq:072}.   
\end{equation}
\textcolor{blue}{Thus, $\mathbf{G}$ contains $2N_I+1$ dominant FTN correlation coefficients and models the finite memory ISI introduced by FTN signaling in the discrete receiver model.}

After  applying DAFT, the received signal in the DAFT-domain is given by
 \begin{equation}
	\scalebox{0.9}{$%
		\begin{aligned}
			\mathbf{r}  = \mathbf{H}_{\rm{eff}}\mathbf{x} + \mathbf{\hat{w}},
		\end{aligned}
		$}
	\label{eq:07}   
\end{equation}
where $\mathbf{H}_{\rm{eff}} = \mathbf{AH}\mathbf{G}\mathbf{A}^{H} $, $\mathbf{\hat{w}} = \mathbf{Aw}$ is the noise vector in the DAFT-domain with the same mean and variance as $\mathbf{w}$. 

\begin{figure}[!t]
\centerline{\includegraphics[width=0.9\columnwidth]{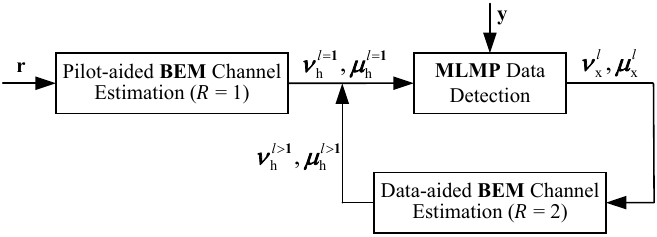}}
	\caption{Joint iterative process of the proposed BEM-MLMP-JCED  algorithm.}
	\label{fig4}
    \vspace{-0.5cm}
\end{figure}
\section{Proposed joint BEM channel estimation and MLMP data detection}
\label{sec:4} 

This section presents the detailed receiver design for the proposed AFDM-FTN system. Specifically, we first derive the input-output relationship of the proposed AFDM-FTN system in Section~\ref{Sec3.0}, which lays the foundation for the subsequent pilot design. We then introduce the detailed BEM model and the proposed MLMP detector. Finally, Section~\ref{sec:3.2} presents the proposed BEM-MLMP-JCED scheme, which consists of the BEM-assisted channel estimation   and its joint iterative integration with the MLMP detector, as illustrated in Fig.~\ref{fig4}. In the BEM-MLMP-JCED framework, the detected data symbols are iteratively refined to enhance the estimation accuracy.
\vspace{-0.5em}
\subsection{Analysis of the Input-output Relationship  for  AFDM-FTN}
\label{Sec3.0}
In contrast to conventional AFDM where multiple paths can be separated with suitable parameters \cite{liChirpParameterSelection2025}, FTN intentionally  introduces ISI.
In the following,  we focus on the input-output relationship of the proposed AFDM-FTN system. 

Note that the FTN coefficient matrix $\mathbf{G}$ in \eqref{eq:072} is circulant and can be written as
\begin{equation}
	\scalebox{0.9}{$%
		\begin{aligned}
			\mathbf{G}= \sum_{k=-N_I}^{N_I}g(k) \mathbf{\Pi}^k,
		\end{aligned}
		$}
	\label{eq:3}   
\end{equation}
which shares a similar structure with that of TD multipath channel matrix.  
Define ${\rm{loc}}_{l_i,k} \triangleq \left(\varphi_i+ 2Nc_1(l_i-k)\right)_N$, and by 
 substituting  \eqref{eq:3} into  $\mathbf{H}_{\rm{eff}}$, the element at the $m$th row and the
$n$th column of $\mathbf{H}_{\rm{eff}}$ is given by
\begin{equation}
	\scalebox{0.9}{$%
		\begin{aligned}
			 H_{\rm{eff}}[m,n]=\frac{1}{N} \sum_{i=0}^{L-1}  \sum_{k=-N_I}^{N_I} h_i\mathcal{T}_{k,i}[m,n]\mathcal{F}_{k,i}[m,n],
		\end{aligned}
		$}
	\label{eq:31}   
\end{equation}
with
\begin{equation}
	\scalebox{0.9}{$%
		\begin{aligned}
			\mathcal{T}_{k,i}[m,n]=g[k]e^{j\frac{2\pi}{N}\left(Nc_2(n^2-m^2)+Nc_1(l_i^2-k^2)-l_in+mk\right)},
		\end{aligned}
		$}
	\label{eq:33}   
\end{equation} 
and
\begin{equation}
	\scalebox{0.9}{$%
		\begin{aligned}
			\mathcal{F}_{k,i}[m,n]&=
			\sum_{p=0}^{N-1}e^{j\frac{2\pi}{N}(n-m- {\rm{loc}}_{l_i,k}- \psi_i )p}\\
			&=\frac{e^{j2\pi(n-m-{\rm{loc}}_{l_i,k}- \psi_i )}-1}{e^{j\frac{2\pi}{N}(n-m- {\rm{loc}}_{l_i,k}- \psi_i )}-1},
		\end{aligned}
		$}
	\label{eq:32}   
\end{equation}
where \eqref{eq:32} models the spreading   caused by Doppler.  
In this paper, similar to existing works \cite{yinDiagonallyReconstructedChannel2024}, \cite{bemaniAffineFrequencyDivision2023},  $c_1$ is chosen as follows 
  \begin{equation}
 	\scalebox{0.9}{$%
 		\begin{aligned}
 			c_1 =  \frac{2(\varphi_{\rm{max}}+N_p)+1}{2N},
 		\end{aligned}
 		$}
 	\label{eq:14}   
 \end{equation}
where $N_p$ is a non-negative integer used to counteract fractional Doppler  and $c_2$ can be set to a rational number much smaller than $\frac{1}{2N}$.

\begin{figure}[t]
	\centering
	\subfigure[\textcolor{blue}{Magnitude distribution from the $m$th row of $\mathbf{H}_{\rm{eff}}$ with integer Doppler.}]{
		\includegraphics[width=1\linewidth]{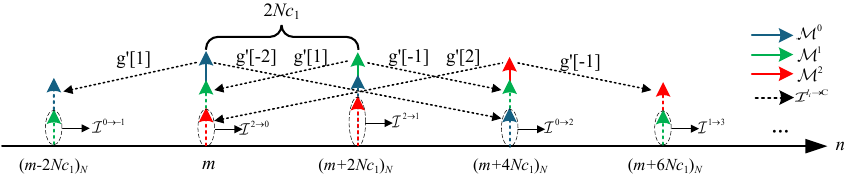}
		\label{fig33:a}}
	\vspace{2pt}
	\subfigure[\textcolor{blue}{Magnitude  distribution from the $m$th row of $\mathbf{H}_{\rm{eff}}$ with fractional Doppler.}]{
		\includegraphics[width=1\linewidth]{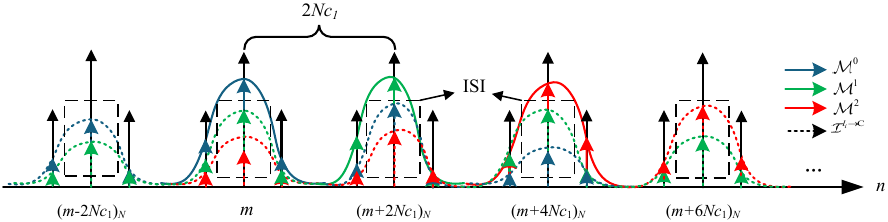}
		\label{fig33:b}
	}
	\caption{\textcolor{blue}{Structural characterization of the effective channel matrix in AFDM-FTN systems.}}
	\label{fig33}
    \vspace{-0.5cm}
\end{figure}
Based on \eqref{eq:3}, the FTN signaling introduces $2N_I+1$ ISI terms. Consequently, in the AFDM-FTN,    additional  $2N_I+1$   components are introduced for each propagation path in the resulting effective channel matrix, as shown in \eqref{eq:31}. In the following, we provide a detailed analysis of the interference structure of the effective channel matrix by considering the cases of integer and fractional Doppler shifts, respectively.


\textit{1) Integer Doppler }: In the case of integer Doppler, i.e., $\psi_i = 0$, it holds that $\mathcal{F}_{k,i}[m,n] = N$ for $n = (m+\mathrm{loc}_{l_i,k})_N$, and $\mathcal{F}_{k,i}[m,n] = 0$ otherwise.  Then, \eqref{eq:31} reduces to
\begin{equation}
    \scalebox{0.9}{$%
		H_{\rm{eff}}[m,n] =
    \begin{cases}
	       \displaystyle \sum\limits_{(k,i) \in \mathcal{D} }h_i\mathcal{T}_{k,i}[m,n], & \!\!\! n = (m+{\rm{loc}}_{l_i,k})_N \\
	       0, & \!\!\! \text{otherwise}
    \end{cases}
    $}
	\label{eq:34},
\end{equation} 
where $ \mathcal{D}  \triangleq \{\, (k,i) \mid l_i - k  \in \mathbb Z, \forall l_i, \forall k \,\}$. 
For integer Doppler, the effective channel matrix $\mathbf{H}_{\rm{eff}}$ therefore has nonzero entries only on a finite number of diagonals.

Compared with the effective channel matrix of the conventional AFDM system, FTN causes multipath components with the same Doppler index to overlap at positions $n = (m+{\rm{loc}}_{l_i,k})_N$.
To illustrate this effect, 
Fig. \ref{fig33:a} shows the magnitude of the $m$th row of $\mathbf{H}_{\rm{eff}}$ for three paths with $l_i= 0, 1, 2$ and $\varphi_i = 0$ for all paths.
\textcolor{blue}{Define $\mathcal{M}^{l_i}$ as the desired component at $n = (m+{\rm{loc}}_{l_i,0})_N$,
and let $\mathcal{I}^{l_i \to {\rm{C}}}$ be  the ISI component introduced by $\mathcal{M}^{l_i}$   in the  $m$th row and $((m+2Nc_1{\rm{C}})_N)$th column   with FTN-ISI  coefficient $g[k]$, where   ${\rm{C}}  \triangleq l_i-k$.}
In the integer Doppler case, we have $\mathcal{M}^{l_i} = |g[0]h_i|$ and $\mathcal{I}^{l_i \to {\rm{C}}} = g'[k]\mathcal{M}^{l_i}$, where $g'[k] = g[k]/g[0]$.
For example, at $m=n$ in Fig. \ref{fig33:a}, the interference consists of $\mathcal{I}^{1 \to 0}$ and $\mathcal{I}^{2 \to 0}$. 
For integer Doppler, the diagonal pattern of $\mathbf{H}_{\rm{eff}}$ can  uniquely determine the Doppler indices. In addition, both the FTN-induced ISI pattern and the corresponding FTN coefficients $g[k]$ are known. Hence, the Doppler shifts can be readily estimated from $\mathbf{H}_{\rm{eff}}$ using conventional linear estimation methods or exhaustive search.


\begin{figure}[t]
\centerline{\includegraphics[width=0.9\columnwidth]{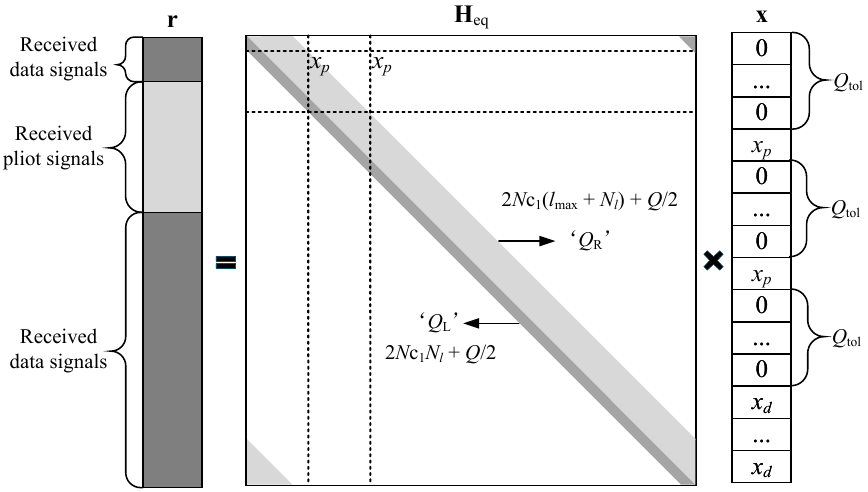}}
	\caption{BEM-assisted channel estimation for the proposed AFDM-FTN system.}
	\label{fig2}
    \vspace{-0.2cm}
\end{figure}

\textit{2) Fractional Doppler }:  In the case of   fractional Doppler,  $\mathcal{M}^{l_i}$ attains its maximum value at $n = (m+{\rm{loc}}_{l_i,0})_N$ and decreases as $n$ moves away from  $m+{\rm{loc}}_{l_i,0}$.
\textcolor{blue}{Fig. \ref{fig33:b} illustrates the magnitude distribution in the $m$th row of the $\mathbf{H}_{\rm{eff}}$ in the presence of fractional Doppler, where the envelope of the Doppler-induced spreading is also shown to highlight the interference dispersion around the dominant FTN-induced components.}
For the AFDM-FTN system, the terms $\mathcal{I}^{l_i \to {\rm{C}}}$ contributed by adjacent paths are summed vectorially around each location $n = (m+{\rm{loc}}_{l_i,k})_N$. 
In this case, \eqref{eq:32} can be rewritten as
\begin{equation}
	\scalebox{0.9}{$%
		\begin{aligned}
			\mathcal{F}_{k,i}[m,n]= \frac{\sin(N\Theta^{m,n}_{k,i})}{\sin(\Theta^{m,n}_{k,i})}e^{j(N-1)\Theta^{m,n}_{k,i}}
		\end{aligned},
		$}
	\label{eq:3212}   
\end{equation}
where $\Theta^{m,n}_{k,i} = \frac{\pi}{N}(n-m-2Nc_1(l_i-k)-\hat{f}_i)$. 
Assuming a sufficiently large guard region to suppress interference between the pilot and data symbols, the  pilot signal received at the index $m$ can be expressed as
\begin{equation}
	\scalebox{0.9}{$%
		\begin{aligned}
			\displaystyle r_p[m] &= \frac{1}{N}\sum\limits_{k,i \in \mathcal{D}}h_i\mathcal{T}_{k,i}[m,n']\frac{\sin(N\Theta^{m,n'}_{k,i})}{\sin(\Theta^{m,n'}_{k,i})}e^{j(N-1)\Theta^{m,n'}_{k,i}} x_p\\
            &+ \hat{w}[m],
		\end{aligned}
		$}
	\label{eq:341}   
\end{equation}
where $x_p$ represents the transmitted pilot symbol with index of $n'$ and $r_p[m]$ is the $m$th received pilot symbol. 
For each pilot observation $r_p[m]$ with known time index $m$
and pilot symbol index $n'$,  $\mathcal{T}_{k,i}[m,n']$ and $x_p$ are known, while $h_i$ and $\hat{f}_i$ are the unknown parameters to be estimated.
Due to the superposition of multiple unknown terms introduced by FTN, more observation samples are required to  achieve reliable estimation, which inevitably increases the pilot overhead. 
In addition, solving the resulting nonlinear system of equations with exponential unknowns is typically ill-conditioned, which further complicates direct parameter estimation based on \eqref{eq:341}.

\vspace{-0.5em}
\subsection{BEM-assisted Channel Estimation}
\label{sec:3-2} 
As discussed above, channel estimation in AFDM-FTN is more challenging than in conventional AFDM, since the parameters $h_i$ and $\hat{f}_i$ are coupled in a nonlinear and highly coupled manner. 
To address this challenge, we propose a BEM-MLMP-JCED receiver. In the following, we first present the fundamentals of the BEM framework and the associated pilot structure, whereas the detailed BEM-MLMP-JCED receiver is described in Section \ref{sec:3.2}. 

By exploiting BEM, the TD channel of the $\ell$th  path at the $t$th time instant can be modeled as
\begin{equation}
	\scalebox{0.9}{$%
		\begin{aligned}
			h(t,\ell) = \sum_{q = 0}^{Q} c_q[\ell]e^{j\omega _qt} + e_{\rm{mod}}[t,\ell],
		\end{aligned}
		$}
	\label{eq:1}   
\end{equation}
where $Q$ denotes the order of BEM, $\omega_q = 2\pi(q-\left \lceil Q/2 \right \rceil )/NR$, $c_q[\ell]$ is the $q$th BEM coefficient of the $\ell$th path, $e_{\rm{mod}}$ is the model error, and $R$ is the resolution parameter ($R\ge 2$ in general \cite{liuNearOptimalBEMOTFS2022}). 
\eqref{eq:1} can be rewritten in matrix form as
\begin{equation}
	\scalebox{0.9}{$%
		\begin{aligned}
			\mathbf{H} = \sum_{q=0}^{Q}{\rm{diag}}\{ \mathbf{u}_q\}\mathbf{C}_q+\mathbf{E}_{\rm{mod}},
		\end{aligned}
		$}
	\label{eq:101}   
\end{equation}
where $\mathbf{u}_q = \{e^{-j2\pi \omega_q n/N},n=0,1,\cdots,N-1\}$ is the BEM basis vector, $\mathbf{C}_q$ is a circulant matrix whose first column is $\mathbf{c}_q=[c_q[0],c_q[1],\cdots,c_q[L-1],0,\cdots,0]^T$, $\mathbf{E}_{\rm{mod}}$ is the matrix form of $e_{\rm{mod}}$. With BEM \textcolor{blue}{and define $ \chi_{k,i,q}(m,n) \triangleq n-m+\omega_q-2Nc_1(l_i-k)$,}  \eqref{eq:32} can be rewritten as
\begin{equation}
	\textcolor{blue}{\scalebox{0.9}{$%
		\begin{aligned}
			\mathcal{F}_{k,i}[m,n]=\sum_{q=0}^{Q}c_q[l_i]\sum_{p=0}^{N-1}e^{j\frac{2\pi}{N}(\chi_{k,i,q}(m,n))p}.
		\end{aligned}
		$}}
	\label{eq:36}   
\end{equation}
We observe that when Doppler is decomposed into multiple fixed frequency components $\omega_q$, $\mathcal{F}_{k,i}[m,n]$  from different paths can be directly merged, such that only the BEM coefficients need to be estimated.  
\textcolor{blue}{This significantly reduces the channel estimation complexity from  $LN$ to $L(Q+1)$.}
Substituting \eqref{eq:36} into \eqref{eq:31} yields
\begin{equation}
	\textcolor{blue}{\scalebox{0.88}{$%
		\begin{aligned}
			 H_{\rm{eff}}[m,n] =\frac{1}{N}\sum_{i=0}^{L-1}\sum_{k=-N_I}^{N_I}\mathcal{T}_{k,i}[m,n]
             \sum_{q=0}^{Q} c_q[l_i]\frac{e^{j2\pi(\chi_{k,i,q}(m,n))}-1}{e^{j\frac{2\pi}{N}(\chi_{k,i,q}(m,n))}-1}
		\end{aligned}.
		$}}
	\label{eq:4}   
\end{equation}
\textcolor{blue}{For each triplet $(i,k,q)$, the corresponding summand is concentrated around the shifted diagonal satisfying $\chi_{k,i,q}(m,n)\approx 0$. Thus, \eqref{eq:4} represents a superposition of shifted diagonal contributions controlled by the FTN shift, path delay, and BEM component, which underlies the subsequent pilot protection and guard length design.}


In the following, we introduce the proposed pilot design based on the interference characteristics.
As shown in Fig. \ref{fig2}, 
the transmitted signal consists of pilot (denoted as $x_p$), data (denoted as $x_d$), and guard subcarriers.
To mitigate the impact of  ISI caused by FTN, guard intervals with a length of  $2Nc_1N_l$ are inserted on both sides of the pilot, where $N_l$ is the FTN-ISI truncation length.
It should be noted that the size of $N_l$ depends on the FTN compression factor. Generally, the smaller the compression factor, the larger the value of $N_l$.
Accordingly, we choose the guard length between the two pilots as
 \begin{equation}
 	\scalebox{0.9}{$%
 		\begin{aligned}
 			Q_{\rm{tol}} = \underbrace{\frac{Q}{2} +2Nc_1N_l }_{Q_{\text{L}}} + \underbrace{\frac{Q}{2}+2Nc_1(l_{\rm{max}} +N_l)}_{Q_{\text{R}}},
 		\end{aligned}
 		$}
 	\label{eq:12}   
 \end{equation}
where $l_{\rm{max}}$ is the maximum delay, and $Q_{\text{L}}$  and $Q_{\text{R}}$  denote the  left and right guard lengths, respectively. 

\textcolor{blue}{Based on the above guard design, the pilot overhead of the proposed AFDM-FTN scheme can be expressed as $\rho_{\rm AFDM\mbox{-}FTN}=(N_{\rm pilot}+Q_{\rm tol})/N$, where $N_{\rm pilot}$ denotes the number of pilot symbols. In contrast, for conventional AFDM without FTN-ISI, the pilot overhead can be written as $\rho_{\rm AFDM}=(N_{\rm pilot}+N_g^{\rm AFDM})/N$, where $N_g^{\rm AFDM}$ is the guard length required to isolate the pilot from delay-Doppler spreading. In particular, when $N_l=0$, the proposed guard design reduces to the conventional AFDM case, yielding $N_g^{\rm AFDM}=Q_{\rm tol}$. Compared with conventional AFDM, the additional overhead of AFDM-FTN mainly comes from the FTN-ISI region included in \(Q_{\rm tol}\). Nevertheless, the proposed BEM-assisted approach reduces the channel estimation problem to the estimation of $L(Q+1)$ BEM coefficients, thereby avoiding excessive pilot observations required for directly estimating the full time-varying channel response.}

\vspace{-0.5em}
\subsection{MLMP-based Data Detection}
\label{sec:3.1} 
Benefiting from the inherent sparsity in the DAFT domain,  message-passing-based low-complexity detectors have been widely adopted in AFDM systems \cite{luoAFDMSCMAPromisingWaveform2024}, \cite{11150613}. However, with FTN signaling, the effective channel $\mathbf{H}_{\rm eff}$ becomes  significantly denser than in conventional AFDM systems, which reduces the efficiency of   message passing schemes when operating directly on $\mathbf{H}_{\rm eff}$. 
Note that $\mathbf{H}_{\rm eff}$ can be factorized into $\mathbf{H}$, $\mathbf{G}$, and $\mathbf{A}$, where $\mathbf{H}$ has a well-defined sparsity pattern determined by $l_{\max}$, $\mathbf{G}$ is a known sparse matrix with only a few nonzero taps, and $\mathbf{A}$ is a unitary transform matrix that supports efficient forward and inverse operations. By exploiting this structured factorization, we propose an efficient  MLMP algorithm for  detection.

Based on the MAP detection, the  marginal distribution of  $x_n$ is given by
\begin{equation}
 	\scalebox{0.9}{$%
 \begin{aligned}  
 p\!\left(x_n | \mathbf{y}\right) 
 \propto 
 \int_{\mathbf{H}, \mathbf{x}\setminus x_n} 
 p\!\left(\mathbf{x}, \mathbf{H} | \mathbf{y}\right).
  	\end{aligned}
    $}
 \label{eq:221}   
\end{equation}
According to Bayes' rule, $p(\mathbf{x}, \mathbf{H} | \mathbf{y})$ can be expressed as
\begin{equation}
	\begin{aligned}  
		p(\mathbf{x}, \mathbf{H} | \mathbf{y})
		\propto 
		p(\mathbf{x}) p(\mathbf{H})  p(\mathbf{y} | \mathbf{x}, \mathbf{H}  ).
	\end{aligned}
	\label{eq:222}   
\end{equation}
In \eqref{eq:222}, $p(\mathbf{x})$ and $p(\mathbf{H})$ are specified by the adopted priors.
To enable efficient message passing,
we exploit the sparse structure of $\mathbf{H}$ induced by the delay taps $\{l_i\}_{i=0}^{L-1}$
and introduce the auxiliary variables $\mathbf{s}=\mathbf{G}\bar{\mathbf{x}}$ and $\bar{\mathbf{x}}=\mathbf{A}^H\mathbf{x}$, which leads to the following factorization of $p(\mathbf{y}| \mathbf{x}, \mathbf{H})$ given by
\begin{equation}
\scalebox{0.9}{$%
	\begin{aligned}  
		p(\mathbf{y}| \mathbf{x}, \mathbf{H})
		\propto  
		\prod_{m,n} &\underbrace{\delta(y_m-\sum_{n\in \mathcal{N}_f(m)}h_{mn}s_n)}_{f_m}\\  &\underbrace{\delta(s_n-\sum_{k=-N_l}^{N_l}g_k \bar{x}_{n-k})}_{\phi_n}
		\underbrace{\delta\left(\bar{x}_n-(\mathbf{A}^H \mathbf{x})_n\right)}_{\mathcal{A}_n},
	\end{aligned}
    $}
	\label{eq:223}   
\end{equation}
where  $h_{mn}$ represents the element in the $m$th row and $n$th column of $\mathbf{H}$.
For each observation $y_m$, $\mathcal{N}_f(m)$ denotes the set of variable node (VN) indices connected to the function node (FN) $f_m$, i.e., $\mathcal{N}_{f}(m) \triangleq  \left\{ n \in \mathbb{Z} | n = m - l_i, 0 \le i \le L-1 \right\}$,
and, conversely, for a given variable index $n$, the set of FN $f_m$ indices connected to the corresponding VN is defined as $\mathcal{N}_{s}(n) \triangleq  \left\{ m \in \mathbb{Z} | m = n + l_i, 0 \le i \le L-1 \right\}$.

\begin{figure}[t]
\centerline{\includegraphics[width=0.9\columnwidth]{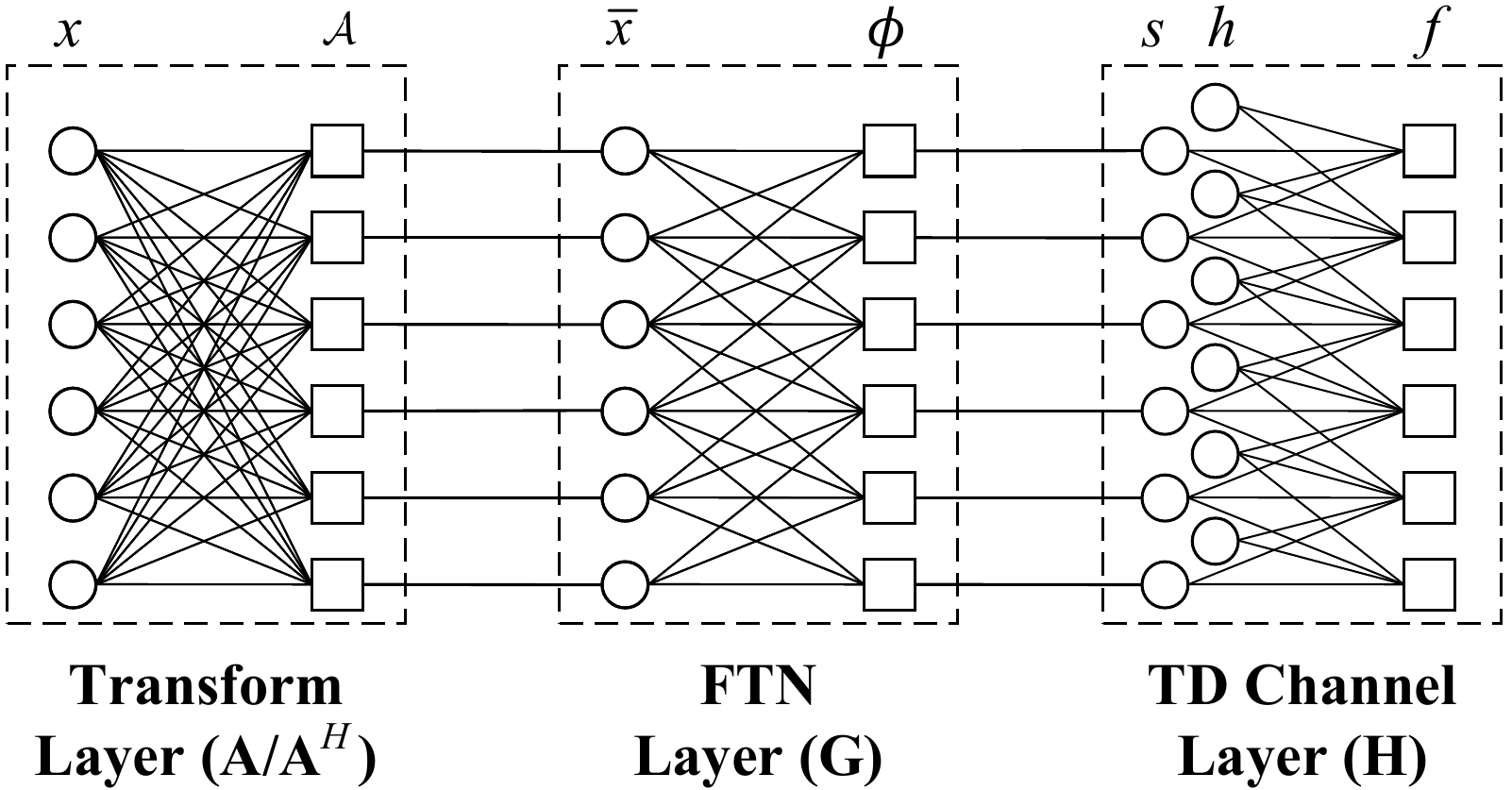}}
	\caption{Message passing process of the MLMP algorithm.}
	\label{fig3}
    \vspace{-0.5cm}
\end{figure}

The decomposition of $p(\mathbf{y}| \mathbf{x}, \mathbf{H})$ naturally leads to a three-layer message passing structure, which can be visualized using a factor graph shown in Fig. \ref{fig3},  where $f, \phi, \mathcal{A}$ are FNs and $h, s, \bar{x}, x$ are VNs.  
In each iteration, messages are updated sequentially across three layers:  the TD channel layer connected through $\mathbf{H}$ updates the messages of $h_{mn}$ and  $s_n$, the FTN layer connected through $\mathbf{G}$ updates the messages between $s_n$ and $\bar{x}_n$, and the transform layer connected through $\mathbf{A} / \mathbf{A}^{H}$ updates the messages between $\bar{x}_n$ and  $x_n$.
Let $\mu_{a}$ and $\nu_{a}$ denote the mean and variance of the marginal message at node $a$, and let $\mu_{a\to b}$ and  $\nu_{a\to b}$  denote the mean and variance of the message propagated from node $a$ to node $b$, respectively.
In what follows, we present the detailed message passing of the proposed MLMP algorithm. 

\textbf{1) TD Channel layer:} 
In this layer, the message passing procedure is decomposed into two components. The first component performs message updates between $f\leftrightarrow h$ based on the current estimate of the channel coefficient $h$. The update rules for $h$ are detailed in Section \ref{sec:3.2}. The second component exploits this updated channel estimate to refine the data messages between $f\leftrightarrow s$. 

\textbf{1.1) Message passing between $f\leftrightarrow h$ :}
Using pilot signals for initial channel estimation, we can obtain the probability distribution   of  channel parameters, i.e., $p(h_{mn})$, which can be modeled as a complex Gaussian distribution with  mean $\mu_{h_{mn}}$ and variance $\nu_{h_{mn}}$. Assume that $f\to h$ has complex Gaussian forms with mean $\mu_{f_m\to h_{mn}}$ and variance $\nu_{f_m\to h_{mn}}$.
Then, based on the message propagation rule \cite{yuanIterativeReceiverDesign2020}, one has 
\begin{equation}
\scalebox{0.9}{$%
	\begin{aligned}  
		\nu_{h_{mn}\to f_m}=\left(\sum_{j\in\mathcal N_s(n)\setminus m}\frac{1}{\nu_{f_{j}\to h_{mn}}}+\frac{1}{\nu_{h_{mn}}}\right)^{-1},
	\end{aligned}
        $}
	\label{eq:224}   
\end{equation}
\begin{equation}
	\scalebox{0.9}{$%
	\begin{aligned}  
		\mu_{h_{mn}\to f_m}=\nu_{h_{mn}\to f_m}\left(\sum_{j\in\mathcal N_s(n)\setminus m}\frac{\mu_{f_{j}\to h_{mn}}}{
			\nu_{f_{j}\to h_{mn}}}+\frac{\mu_{h_{mn}}}{\nu_{h_{mn}}}\right).
	\end{aligned}
    $}
	\label{eq:21}   
\end{equation}
In the factor graph representation, the TD received sample associated with FN $f_m$ can be decomposed as
\begin{equation}
	\scalebox{0.9}{$%
	\begin{aligned}  
		y_m = \underbrace{h_{mn}s_n}_{\text{current edge}} + \underbrace{\sum_{j\in\mathcal N_f(m)\setminus n}h_{mj}s_j}_{\text{interference edge}} + \, w_m.
	\end{aligned}
            $}
	\label{eq:21110}   
\end{equation}
By modeling the remaining edges as interference and approximating them with a complex Gaussian distribution, the mean and variance of these interference components are respectively given by
\begin{subequations}
	\begin{equation}
    \scalebox{0.9}{$%
    \begin{aligned} 
    \mu_{Ih,n} &=\sum_{j\in\mathcal N_f(m)\setminus n} \mu_{h_{mn}\to f_j} \mu_{s_{j}\to f_m},
    \end{aligned}
    $}
    \label{eq:211b} 
    \end{equation}
	\begin{equation}
        \scalebox{0.9}{$%
            \begin{aligned} 
               \sigma_{Ih,n}^2 &= \sum_{j\in\mathcal N_f(m)\setminus n}|\mu_{s_{j}\to f_m}|^2\,\nu_{h_{mn}\to f_j} \\
		&+ \sum_{j\in\mathcal N_f(m)}\nu_{s_{j}\to f_m}(|\mu_{h_{mn}\to f_j}|^2+\nu_{h_{mn}\to f_j}).
            \end{aligned}
        $}
    \label{eq:21a}  
    \end{equation} 
    \label{eq:211a}%
\end{subequations}
By combining the interference and noise into an effective noise term and subtracting the interference mean, we obtain the equivalent observation as
\begin{equation}
	\scalebox{0.9}{$%
		\begin{aligned}  
			\tilde{y}_m  =y_m-\mu_{Ih,n} =s_nh_{mn}+n_m, 
		\end{aligned}
	$}
	\label{eq:2110}   
\end{equation}
\textcolor{blue}{where $n_m \sim \mathcal{CN}\left(0, \sigma^2+\sigma_{Ih,n}^2\right)$. During belief propagation, the message from $f_m$ to $h_{mn}$ is obtained by marginalizing over $s_n$, i.e.,}
\begin{equation}
	\scalebox{0.9}{$%
		\begin{aligned} 
			\textcolor{blue}{m_{f_m\rightarrow h_{mn}}(h_{mn}) \propto \int p(\tilde{y}_m|h_{mn},s_n)m_{s_n\rightarrow f_m}(s_n){\rm{d}}s_n,}
		\end{aligned}
$}
\label{eq:21103}   
\end{equation}
\textcolor{blue}{where $m_{s_n\rightarrow f_m}(s_n)$ denotes the incoming message from VN $s_n$ to FN $f_m$, which is approximated as $m_{s_n\rightarrow f_m}(s_n)=\mathcal{CN}\left(s_n;\mu_{s_n\rightarrow f_m},\nu_{s_n\rightarrow f_m}\right)$. According to the equivalent scalar model in \eqref{eq:2110}, the conditional likelihood is given by}
\begin{equation}
	\scalebox{0.9}{$%
		\begin{aligned} 
			\textcolor{blue}{p(\tilde{y}_m|h_{mn},s_n) \propto\exp\left(-\frac{|\tilde{y}_m-s_n h_{mn}|^2}{\sigma^2+\sigma^2_{I_{h,n}}}\right).}
		\end{aligned}
	$}
\label{eq:21104}   
\end{equation}
\textcolor{blue}{Since the incoming message of $s_n$ is approximated as $s_n\sim\mathcal{CN}(\mu_{s_n\rightarrow f_m},\nu_{s_n\rightarrow f_m})$, we apply Gaussian projection to the bilinear term $s_n h_{mn}$. Specifically,}
\begin{equation}
	\scalebox{0.9}{$%
	\textcolor{blue}{
		\begin{aligned} 
		&\mathbb{E}_{s_n} \left[|\tilde{y}_m-s_n h_{mn}|^2\right]\\
		&\quad=|\tilde{y}_m|^2-\tilde{y}_m^*h_{mn}\mu_{s_n\rightarrow f_m}-\tilde{y}_m h_{mn}^*\mu^*_{s_n\rightarrow f_m}\\
		&\qquad+|h_{mn}|^2\left(|\mu_{s_n\rightarrow f_m}|^2+\nu_{s_n\rightarrow f_m}\right)\\
		&\quad=\left(|\mu_{s_n\rightarrow f_m}|^2+\nu_{s_n\rightarrow f_m}\right)\left|h_{mn}-\frac{\tilde{y}_m\mu^*_{s_n\rightarrow f_m}}{|\mu_{s_n\rightarrow f_m}|^2+\nu_{s_n\rightarrow f_m}}\right|^2\\
		&\qquad+
		\frac{|\tilde{y}_m|^2\nu_{s_n\rightarrow f_m}}
		{|\mu_{s_n\rightarrow f_m}|^2+\nu_{s_n\rightarrow f_m}}
		\end{aligned}.}
	$}
\label{eq:21101}   
\end{equation}
\textcolor{blue}{Therefore, the message from $f_m$ to $h_{mn}$ can be approximated as}
\begin{equation}
	\scalebox{0.9}{$%
		\begin{aligned}
			\textcolor{blue}{m_{f_m\rightarrow h_{mn}}(h_{mn})\propto\exp\left(-\frac{|h_{mn}-\mu_{f_m\rightarrow h_{mn}}|^2}{\nu_{f_m\rightarrow h_{mn}}}\right),}
		\end{aligned}
	$}
\label{eq:21102}   
\end{equation}
\textcolor{blue}{where  the mean and variance are respectively given by}
\begin{equation}
	\begin{aligned}  
		\mu_{f_m\to h_{mn}}=\frac{(y_m-\mu_{Ih,n})\mu^{*}_{s_n\to f_m}}{|\mu_{s_{n}\to f_m}|^2 + \nu_{s_{n}\to f_m}},
	\end{aligned}
	\label{eq:2111}   
\end{equation}
\textcolor{blue}{and}
\begin{equation}
	\begin{aligned}  
	\nu_{f_m\to h_{mn}} &= \frac{\sigma^2+\sigma_{Ih,n}^2}{|\mu_{s_{n}\to f_m}|^2 + \nu_{s_{n}\to f_m}}.
	\end{aligned}
	\label{eq:2112}   
\end{equation} 

\textbf{1.2) Message passing between $f\leftrightarrow s$ :}
\textcolor{blue}{The messages from $f_m$ to $s_n$ can be derived in an analogous manner by treating $s_n$ as the target variable and $h_{mn}$ as the Gaussian-distributed coefficient. Accordingly, the variance and mean   of  $f\to s$ are respectively given by}
\begin{equation}
	\begin{aligned}  
		\nu_{f_m\to s_n} &= \frac{\sigma^2+\sigma_{Is,n}^2}{|\mu_{h_{mn}\to f_m}|^2 + \nu_{h_{mn}\to f_m}},
	\end{aligned}
	\label{eq:2212}   
\end{equation}
\begin{equation}
	\begin{aligned}  
		\mu_{f_m\to s_n}=\frac{(y_m-\mu_{Is,n})\mu^{*}_{h_{mn}\to f_m}}{|\mu_{h_{mn}\to f_m}|^2 + \nu_{h_{mn}\to f_m}},
	\end{aligned}
	\label{eq:2211}   
\end{equation}
where
\begin{subequations}
	\begin{equation}
    \scalebox{0.9}{$%
    \begin{aligned} 
    \mu_{Is,n}&=\sum_{j\in \mathcal N_f(m)\setminus n} \mu_{s_{j}\to f_m} \mu_{h_{mn}\to f_j},
    \end{aligned}
    $}
    \label{eq:221a} 
    \end{equation}
	\begin{equation}
        \scalebox{0.9}{$%
            \begin{aligned} 
               \sigma_{Is,n}^2 &= \sum_{j\in \mathcal N_f(m)\setminus n}|\mu_{h_{mn}\to f_j}|^2\,\nu_{s_{j}\to f_m} \\
		& + \sum_{j\in \mathcal N_f(m)}\nu_{s_{j}\to f_m}(|\mu_{h_{mn}\to f_j}|^2+\nu_{h_{mn}\to f_j}).
            \end{aligned}
        $}
    \label{eq:221b}  
    \end{equation} 
    \label{eq:2213}%
\end{subequations}
According to the Gaussian message passing rules, the variance and  mean  from VN $s_n$ to FN $f_m$ are respectively given by
\begin{equation}
    \scalebox{0.9}{$%
	\begin{aligned}  
			\nu_{s_n\to f_m}= \left(\sum_{j\in\mathcal N_s(n)\setminus m}\frac{1}{\nu_{f_{j}\to s_n}}+\frac{1}{\nu_{\phi_n\to s_n}}\right)^{-1},
	\end{aligned}
    $}
	\label{eq:231}   
\end{equation}
\begin{equation}
    \scalebox{0.9}{$%
		\begin{aligned}  
			\mu_{s_n\to f_m}=\nu_{s_n\to f_m}\left(\sum_{j\in\mathcal N_s(n)\setminus m}\frac{\mu_{f_{j}\to s_n}}{
				\nu_{f_{j}\to s_n}}+\frac{\mu_{\phi_n\to s_n}}{\nu_{\phi_n\to s_n}}\right),
		\end{aligned}
            $}
		\label{eq:232} 
\end{equation}
where the message of $\phi_n\to s_n$ arises from message passing across layers and is obtained from the FTN layer message, with its  variance and mean given by
\begin{equation}
\scalebox{0.9}{$%
	\begin{aligned}  
		\nu_{\phi_n\to s_n}=\sum_{j=-N_l}^{N_l}|g_{n-j}|^2\,\nu_{\bar{x}_j\to \phi_n},
	\end{aligned}
    $}
	\label{eq:25}   
\end{equation}
\begin{equation}
\scalebox{0.9}{$%
	\begin{aligned}  
		\mu_{\phi_n\to s_n}=\sum_{j=-N_l}^{N_l}g_{n-j}\mu_{\bar{x}_j\to \phi_n}.
	\end{aligned}
    $}
	\label{eq:24}   
\end{equation}

\textbf{2) FTN  layer:} 
For a specified FTN compression factor $\alpha$, the FTN coefficient matrix $\mathbf G$ remain fixed and exhibit inherent sparsity, thereby enabling efficient message passing. 
Define  $\mathcal N_\phi(j)\triangleq \{n \in \mathbb{Z} \mid j-N_l \le n \le j+N_l \} $ represent the set of all $\phi_n$ connected to $\bar{x}_j$.
Note that $s_n$ can be written in a form similar to \eqref{eq:21110}, then the interference of adjacent symbols at the $j$th VN $\bar{x}_j$ is approximately in complex Gaussian form, and its mean and variance can be respectively calculated as
\begin{subequations}
	\begin{equation}
    \scalebox{0.9}{$%
    \begin{aligned} 
    \hat \mu_{\bar{x}_j}=\sum_{k=-N_l,k\neq j}^{N_l} g_{n-k}\,\mu_{\bar{x}_k\to \phi_n},
    \end{aligned}
    $}
    \label{eq:26a} 
    \end{equation}
	\begin{equation}
        \scalebox{0.9}{$%
            \begin{aligned} 
                \hat \nu_{\bar{x}_j}=\sum_{k=-N_l,k\neq j}^{N_l}|g_{n-k}|^2\nu_{\bar{x}_k\to \phi_n}.
            \end{aligned}
        $}
    \label{eq:26b}  
    \end{equation} 
    \label{eq:226b}%
\end{subequations}
\textcolor{blue}{For the message update from $\phi_n$ to $\bar{x}_j$, the FTN constraint is first rewritten as}
\begin{equation}
	\scalebox{0.9}{$%
		\begin{aligned} 
			\textcolor{blue}{s_n=g_{n-j}\bar{x}_j+\hat{\bar{x}}_j,}
        \end{aligned}
	$}
\label{eq:271}
\end{equation}
\textcolor{blue}{where $\hat{\bar{x}}_j$ denotes the aggregate contribution of the neighboring symbols excluding $\bar{x}_j$. Under the Gaussian message passing approximation, $\hat{\bar{x}}_j$ is modeled as equivalent Gaussian interference with mean and variance given in \eqref{eq:226b}, based on the mutually independent Gaussian extrinsic messages of the neighboring symbols. Incorporating the incoming Gaussian belief of $s_n$, the local FTN constraint yields $\bar{x}_j=(s_n-\hat{\bar{x}}_j)/g_{n-j}$. Hence, the message $\phi_n\rightarrow\bar{x}_j$ is obtained via a standard linear Gaussian transformation, with variance and mean respectively given by}
\begin{equation}
        \scalebox{0.9}{$%
            \begin{aligned} 
    \nu_{\phi_n\to \bar{x}_j}=\frac{\nu_{s_n\to \phi_n}+\hat \nu_{\bar{x}_j}}{|g_{n-j}|^2},  \quad 
    \mu_{\phi_n\to \bar{x}_j}=\frac{\mu_{s_n\to \phi_n}-\hat {\mu}_{\bar{x}_j}}{g_{n-j}},
        \end{aligned}
    $}
    \label{eq:27}
\end{equation}
where  the mean and variance of $s_n \to \phi_n$ are obtained from the TD channel layer message, i.e.,
\begin{equation}
    \scalebox{0.9}{$%
	\begin{aligned}  
		\nu_{s_n\to \phi_n}= \left(\sum_{j\in\mathcal N_s(n)}\frac{1}{\nu_{f_j\to s_n}}\right)^{-1}, 
	\end{aligned}
    $}
	\label{eq:233}   
\end{equation}
\begin{equation}
    \scalebox{0.9}{$%
	\begin{aligned}  
		\mu_{s_n\to \phi_n}=\nu_{s_n\to \phi_n}\left(\sum_{j\in\mathcal N_s(n)}\frac{\mu_{f_j\to s_n}}{
			\nu_{f_j\to s_n}}\right).
	\end{aligned}
        $}
	\label{eq:234}   
\end{equation}
Then, we obtain the mean and variance from VN $\bar{x}_j$ to FN $\phi_n$ respectively as
\begin{equation}
\scalebox{0.9}{$%
	\begin{aligned}  
\nu_{\bar{x}_j\to \phi_n}=\left(\sum_{n'\in\mathcal N_\phi(j)\setminus n}\frac{1}{\nu_{\phi_{n'}\to \bar{x}_j}}+\frac{1}{\nu_{\mathcal {A}_j\to \bar{x}_j}}\right)^{-1} ,
	\end{aligned}
        $}
	\label{eq:210}   
\end{equation}
\begin{equation}
    \scalebox{0.9}{$%
	\begin{aligned}  
		\mu_{\bar{x}_j\to \phi_n}=\nu_{\bar{x}_j\to \phi_n}\left(\sum_{n'\in\mathcal N_\phi(j)\setminus n}\frac{\mu_{\phi_{n'}\to \bar{x}_j}}{\nu_{\phi_{n'}\to \bar{x}_j}}+\frac{\mu_{\mathcal {A}_j\to \bar{x}_j}}{\nu_{\mathcal {A}_j\to \bar{x}_j}}\right).
	\end{aligned}
    $}
	\label{eq:211}  
\end{equation}
Finally, the cross-layer message  from the FTN layer to the transform layer, i.e.,  $\bar{x}_j\to \mathcal A$,  can be obtained  by
\begin{equation}
    \scalebox{0.9}{$%
	\begin{aligned}  
		\nu_{\bar{x}_j\to \mathcal {A}_j}=\left(\sum_{n\in\mathcal N_\phi(j)}\frac{1}{\nu_{\phi_n\to \bar{x}_j}}\right)^{-1},
	\end{aligned}
    $}
	\label{eq:212}   
\end{equation}
\begin{equation}
    \scalebox{0.9}{$%
	\begin{aligned}  
		\mu_{\bar{x}_j\to \mathcal {A}_j}=\nu_{\bar{x}_j\to \mathcal {A}_j}\left(\sum_{n\in\mathcal N_\phi(j)}\frac{\mu_{\phi_n\to \bar{x}_j}}{\nu_{\phi_n\to \bar{x}_j}}\right).
	\end{aligned}
    $}
	\label{eq:213}   
\end{equation}	
By collecting the mean and variance of $N$ nodes,  we obtain ${\bm{\mu}}_{{\mathbf{\bar x}}\to \mathcal A} = [\mu_{{\bar x_0}\to \mathcal A_0}, \mu_{{\bar x_1}\to \mathcal A_1},\cdots, \mu_{{\bar x_{N-1}}\to \mathcal A_{N-1}}]^T$ and ${\bm{\nu}}_{{\mathbf{\bar x}}\to \mathcal A} = [\nu_{{\bar x_0}\to \mathcal A_0}, \nu_{{\bar x_1}\to \mathcal A_1},\cdots, \nu_{{\bar x_{N-1}}\to \mathcal A_{N-1}}]^T$. Then,   ${\bm{\mu}}_{{\mathbf{\bar x}}\to \mathcal A}$ and ${\bm{\nu}}_{{\mathbf{\bar x}}\to \mathcal A}$ are forwarded   to the transmission layer for message updates.

\textbf{3) Transform layer:} 
Denote  ${\bm{\nu}}_{\mathcal A\to \mathbf{x}} = [\nu_{\mathcal {A}_0\to x_0}, \nu_{\mathcal {A}_1\to x_1}, \cdots, \nu_{\mathcal {A}_{N-1}\to x_{N-1}}]^T$ and ${\bm{\mu}}_{\mathcal A\to \mathbf{x}} = [\mu_{\mathcal {A}_0\to x_0}, \mu_{\mathcal {A}_1\to x_1}, \cdots, \mu_{\mathcal {A}_{N-1}\to x_{N-1}}]^T$ as the variance and mean vectors of the messages from $\mathcal{A}_n$ to $ x_n$, respectively. By exploiting the unitary property of the AFDM transform matrix $\mathbf{A}$, we can efficiently perform message passing between  $\bar x$ and $ x$.   One has 
\begin{equation}
    \scalebox{0.9}{$%
	\begin{aligned}  
 		{\bm{\nu}}_{\mathcal A\to \mathbf{x}}=|\mathbf{A}|^{2}{\bm{\nu}}_{{\mathbf{\bar x}}\to \mathcal A},\quad {\bm{\mu}}_{\mathcal A\to \mathbf{x}}= \mathbf{A}{\bm{\mu}}_{{\mathbf{\bar x}}\to \mathcal A}.
	\end{aligned}
        $}
	\label{eq:214}   
\end{equation}	
The posteriori distribution of  DAFT-domain symbol $x_n$ is
\begin{equation}
    \scalebox{0.9}{$%
	\begin{aligned}  
		 P(x_n = \xi) \propto p(x_n) \exp\left(-\frac{|\xi-\mu_{\mathcal {A}_n\to x_n}|^2}{\nu_{\mathcal {A}_n\to x_n}}\right), \xi\in\mathbb{M},
	\end{aligned}
     $}
	\label{eq:216}   
\end{equation}	
where $p(x_n)$ denotes the \textit {a priori} probability of $x_n$ which is   assumed to be equiprobable. 
Then,  the \textit {a posterior} probability $ P(x_n = \xi)$ is projected into a complex Gaussian distribution $\mathcal{CN}(\mu_{x_n}^{\rm{post}},\nu_{x_n}^{\rm{post}})$ with
\begin{equation}
	\begin{aligned}  
	\mu_{x_n}^{\rm{post}}=\sum_{\xi\in\mathbb{M}}  \xi P(x_n = \xi), 
	\end{aligned}
	\label{eq:217}   
\end{equation}
\begin{equation}
    \scalebox{0.9}{$%
	\begin{aligned}  
		\nu_{x_n}^{\rm{post}}=\sum_{\xi \in\mathbb{M}} |\xi-\mu_{x_n}^{\rm{post}}|^2P(x_n = \xi).
	\end{aligned}
     $}
	\label{eq:218}   
\end{equation}	
In order to further improve the performance and convergence stability, we introduce a damping factor $\gamma\in (0,1]$ and update the variance and mean as
\begin{equation}
    \scalebox{0.9}{$%
	\begin{aligned}  
		\nu_{x_n}^{l+1}=\left(\frac{1-\gamma}{\nu_{x_n}^{l}} + \frac{\gamma}{\nu_{x_n}^{\rm{post}}}\right)^{-1}, \mu_{x_n}^{l+1}=(1-\gamma)\mu_{x_n}^{l} + \gamma\mu_{x_n}^{\rm{post}},
	\end{aligned}
    $}
	\label{eq:2190}   
\end{equation}	
respectively, where  the superscript $l$ denotes the $l$th iteration. Denote ${\bm{\mu}}_{\rm{x}} = \left[\mu_{x_0},\mu_{x_1},\cdots,\mu_{x_{N-1}} \right]^T$ and ${\bm{\nu}}_{\rm{x}} = \left[\nu_{x_0},\nu_{x_1},\cdots,\nu_{x_{N-1}} \right]^T$. Then,    ${\bm{\mu}}_{\rm{x}}$ and  ${\bm{\nu}}_{\rm{x}}$ serve as pseudo-pilots for the proposed BEM-assisted channel estimation. 

Finally,  the cross layer message $\mathcal A\to {\mathbf{\bar x}}$ from the transform layer to the FTN layer  can be obtained through IDAFT. We denote ${\bm{\nu}}_{\mathcal A\to {\mathbf{\bar x}}} = [\nu_{\mathcal {A}_0\to \bar{x}_0},\nu_{\mathcal {A}_1\to \bar{x}_1},\cdots, \nu_{\mathcal {A}_{N-1}\to \bar{x}_{N-1}}]^T$ and  ${\bm{\mu}}_{\mathcal A\to {\mathbf{\bar x}}} = [\mu_{\mathcal {A}_0\to \bar{x}_0},\mu_{\mathcal {A}_1\to \bar{x}_1},\cdots, \mu_{\mathcal {A}_{N-1}\to \bar{x}_{N-1}}]^T$ as the variance and mean vectors of the messages from $\mathcal{A}_n$ to $ \bar{x}_n$, respectively,  which are given by
\begin{equation}
    \scalebox{0.9}{$%
	\begin{aligned}  
		{\bm{\nu}}_{\mathcal A\to {\mathbf{\bar x}}}=(|\mathbf{A}|^{2})^{T}{\bm{\nu}}_{\rm{x}},\quad {\bm{\mu}}_{\mathcal A\to {\mathbf{\bar x}}}=\mathbf{A}^{H}{\bm{\mu}}_{\rm{x}}.
	\end{aligned}
    $}
	\label{eq:215}   
\end{equation}	
It can be observed that \eqref{eq:215} provides extrinsic information for \eqref{eq:210} and \eqref{eq:211}, thereby enabling reciprocal message updates across the different layers.
\subsection{The Proposed BEM-MLMP-JCED Algorithm}
\label{sec:3.2} 
This subsection introduces the proposed BEM-MLMP-JCED algorithm, which consists of  a BEM-assisted channel estimation scheme and its joint iterative integration with the MLMP detector presented in Section \ref{sec:3.1}. As described in the TD channel layer, the initial channel estimates obtained from the BEM model are utilized as \textit{a priori} information for the MLMP detector. After data detection within the MLMP framework, the resulting soft estimates of $\mathbf{x}$ from the transform layer are further exploited to refine the channel estimation. These two processes are executed iteratively until convergence is achieved.



Substituting \eqref{eq:101} into \eqref{eq:07}, the received signal with BEM can be rewritten as
 \begin{equation}
	\scalebox{0.9}{$%
		\begin{aligned}
			\mathbf{r} &= \sum_{q=0}^Q\mathbf{A}{\rm{diag}}\{\mathbf{u}_q\}\mathbf{C}_q\mathbf{G}\mathbf{A}^H\mathbf{x} + \mathbf{\hat e}_{\rm{mod}}+ \mathbf{\hat{w}}\\
			&\overset{\text{(i)}}{=} \sum_{q=0}^Q\mathbf{A}{\rm{diag}}\{\mathbf{u}_q\}\mathbf{G}\mathbf{F}^H{\rm{diag}}\{\mathbf{F}\mathbf{A}^H\mathbf{x}\}\mathbf{F}\mathbf{c}_q + \mathbf{\hat e}_{\rm{mod}}+ \mathbf{\hat{w}},\\
		\end{aligned}
		$}
	\label{eq:51}   
\end{equation}
where 
$\mathbf{x} = \mathbf{x}_{\rm{p}} + \mathbf{x}_{\rm{d}}$, with $\mathbf{x}_{\rm{p}}$  and $\mathbf{x}_{\rm{d}}$ denoting pilot and data components, respectively, 
 and $\mathbf{\hat e}_{\rm{mod}} = \mathbf{AE}_{\rm{mod}}\mathbf{G}\mathbf{A}^H\mathbf{x}$ represents the interference vector generated by the model error.
$(\text{i})$ is obtained by using $\mathbf{C}_q\mathbf{G} = \mathbf{G}\mathbf{C}_q$ and $\mathbf{C}_q\mathbf{\bar x} = {\rm{diag}}\{\mathbf{F}\mathbf{\bar x}\}\mathbf{c}_q$ as both $\mathbf{C}_q$ and $\mathbf{G}$ are circulant matrices \cite{liuNearOptimalBEMOTFS2022}.

In the  $l$th iteration, the signal estimate is expressed as $\mathbf{\tilde x}^l = \mathbf{x}_{\rm{p}} + {\bm{\mu}}^l_{\rm{x}}$. Define $\mathbf{c} = [\mathbf{c}_0[0],\mathbf{c}_0[1],\dots,\mathbf{c}_0[L-1],\dots,\mathbf{c}_Q[L-1]]^T$ as the BEM coefficient vector. Based on the LMMSE criterion, the mean and variance of  $\mathbf{c}$ are respectively given by
\begin{equation}
    \scalebox{0.9}{$%
	\begin{aligned}  
		{\bm{\nu}}^{l}_c=\left(({\bm{\bar \nu}}_c)^{-1} + \mathbf{\tilde D}^H\mathbf{C}^{-1}_{\rm{w}}\mathbf{\tilde D}\right)^{-1},
	\end{aligned}
    $}
	\label{eq:512}   
\end{equation}
\begin{equation}
    \scalebox{0.9}{$%
	\begin{aligned}  
		{\bm{ \mu}}^{l}_c={\bm{\nu}}^{l}_c\left(({\bm{\bar \nu}}_c)^{-1}{\bm{\bar \mu}}_c + \mathbf{\tilde D}^H\mathbf{C}^{-1}_{\rm{w}}\mathbf{r}\right),
	\end{aligned}
    $}
	\label{eq:513}   
\end{equation}
with
\begin{subequations}
	\begin{equation}
        \scalebox{0.9}{$%
	\begin{aligned} 
    \mathbf{\tilde D} =[\mathbf{\tilde d}_{0,0},\dots,\mathbf{\tilde d}_{0,L-1},\mathbf{\tilde d}_{1,0},\dots,\mathbf{\tilde d}_{Q,L-1}],
    	\end{aligned}
    $}
    \label{eq:514a} 
    \end{equation}
	\begin{equation}
        \scalebox{0.9}{$%
	       \begin{aligned} 
            \mathbf{\tilde d}_{q,i} = \mathbf{A}{\rm{diag}}\{\mathbf{u}_q\}\mathbf{F}^H{\rm{diag}}\{\mathbf{F}\mathbf{A}^H\mathbf{\tilde x}^l\}\mathbf{F}(:,i),
        \end{aligned}
        $}
    \label{eq:514b}  
    \end{equation} 
	\begin{equation}
        \scalebox{0.9}{$%
	       \begin{aligned} 
        \mathbf{C}_{\rm{w}} = \mathbf{\Sigma}_{\rm{w}}+{\rm{diag}}\{{\bm{\nu}}^l_{\rm{x}}\}, 
                \end{aligned}
        $}
        \label{eq:514c} 
    \end{equation} 
    \label{eq:514}%
\end{subequations}
where ${\bm{\bar \mu}}_c$ and ${\bm{\bar \nu}}_c$ are the \textit {a priori} mean and variance of $\mathbf c$,  respectively, and $\mathbf{C}_{\rm{w}}$ is the equivalent noise variance with $\mathbf{\Sigma}_{\rm{w}} = \sigma^2 \mathbf{I}$.

The estimated value of $\mathbf c$ is used for reconstructing the TD channel. According to \eqref{eq:1}, the channel response $\mathbf{h}=[h(0,0),\cdots,h(0,L-1),h(1,0),\cdots,h(N-1,L-1)]^T$ is given by
\begin{equation}
	\scalebox{0.9}{$%
		\begin{aligned}
			\mathbf h=\mathbf{\tilde U}\mathbf c+\mathbf {{e}}_{\rm{mod}},
	\end{aligned}
		$}
\label{eq:802}   
\end{equation}
where $\mathbf{\tilde U} \triangleq \mathbf U \otimes \mathbf I_{L}$ with $\mathbf{U} = [\mathbf{u}_0^T,\mathbf{u}_1^T,\cdots,\mathbf{u}_Q^T]$ is the BEM basis matrix, and $\mathbf{{e}}_{\rm{mod}} = [{e}_{\rm{mod}}(0,0),\cdots,{e}_{\rm{mod}}(0,L-1),{e}_{\rm{mod}}(1,0),\cdots,{e}_{\rm{mod}}(N-1,L-1)]^T$ corresponds to the modeling error.

Denote  ${\bm{\nu}}_{\rm{h}} = [{\nu}_{h_{0,N-l_{\max}+1}},\cdots,{\nu}_{h_{0,0}},\cdots,{\nu}_{h_{N-1,N-1}}]^T$ and ${\bm{\mu}}_{\rm{h}} = [{\mu}_{h_{0,N-l_{\max}+1}},\cdots,{\mu}_{h_{0,0}},\cdots,{\mu}_{h_{N-1,N-1}}]^T$ as the variance and mean vectors of $\mathbf h$, respectively, which can be computed as
\begin{equation}
	\scalebox{0.9}{$%
	\begin{aligned}  
		{\bm{\nu}}^l_{\rm{h}}={\rm{diag}}\{\mathbf{\tilde U}{\bm{\nu}}^{l}_c\mathbf{\tilde U}^H\},\quad
		{\bm{\mu}}^l_{\rm{h}} =\mathbf{\tilde U}{\bm{\mu}}^{l}_c.
	\end{aligned}
    	$}
	\label{eq:516}   
\end{equation}
Finally,   \eqref{eq:516} is used to update ${\nu}_{h_{mn}}$ in \eqref{eq:224} and ${\mu}_{h_{mn}}$ in \eqref{eq:21}.  The channel estimation and MLMP detection are executed iteratively until convergence is achieved. In the following, we summarize the overall BEM-MLMP-JCED algorithm.


\textbf{Algorithmic Summary:}
The detailed procedure of the proposed BEM-MLMP-JCED algorithm is summarized in Algorithm~\ref{alg1}, where $I_{\rm{max}}$ denotes the maximum number of iterations. Specifically, the proposed BEM-MLMP-JCED algorithm sequentially performs both intra-layer and cross-layer message passing across the TD channel, FTN, and transform layers. \textcolor{blue}{Within the TD channel layer, the messages associated with the channel response $\mathbf{h}$ are obtained using the proposed BEM-assisted estimator.} In the transform layer, the soft estimates of $\mathbf{x}$ are fed back to the channel estimator to further enhance its estimation accuracy, thereby forming a joint iterative process. 
The algorithm terminates when the changes in ${\bm{\mu}}^{l}_{\rm{h}}$ and ${\bm{\mu}}^{l}_{\rm{x}}$ between two consecutive iterations are both below the predefined threshold $\epsilon$, or when the maximum number of iterations is reached.
\begin{algorithm}[t]
	\caption{Proposed BEM-MLMP-JCED Algorithm}
	\label{alg1}
	\begin{algorithmic}[1]
		\REQUIRE  
		$\mathbf{y}, \mathbf{r}, \mathbf{G}, \mathbf{A}, \sigma^2, Q$, $I_{\rm{max}}$, and $\epsilon$. 
		\STATE {Initialize the mean and variance of all messages, typically $\mu=0$ and $\nu=1$; the initial channel estimation is performed with  $\mathbf{\tilde x} = \mathbf{x}_{\rm{p}}$ and $R=1$ by \eqref{eq:512}, \eqref{eq:513}, \eqref{eq:514}, and \eqref{eq:516}}.
		\WHILE { $l < I_{\rm{max}}$}
		\STATE { \textbf{\% Update TD channel layer based on  FN  $f$}}
		\STATE {Update  $\nu_{h_{mn}\to f_m}$ and $\mu_{h_{mn}\to f_m}$ by \eqref{eq:224}, \eqref{eq:21}.}
		\STATE {Update  $\nu_{ f_m\to h_{mn}}$ and $\mu_{ f_m\to h_{mn}}$ by \eqref{eq:211a}, \eqref{eq:2111}, \eqref{eq:2112}}.	
		\STATE {Update $\nu_{f_m\to s_n}$ and $\mu_{f_m\to s_n}$ by \eqref{eq:2212}, \eqref{eq:2211}}, \eqref{eq:2213}.
		\STATE {Update  $\nu_{s_n\to f_m}$ and $\mu_{s_n\to f_m}$  by \eqref{eq:231}}, \eqref{eq:232}.
		\STATE {Update  $\nu_{\phi_n\to s_n}$ and $\mu_{\phi_n\to s_n}$  by \eqref{eq:25},  \eqref{eq:24}}.
        
		\STATE { \textbf{\% Update FTN layer based on  FN $\phi$}}
		\STATE {Update  $\nu_{s_n\to \phi_n}$ and $\mu_{s_n\to \phi_n}$  by \eqref{eq:233},  \eqref{eq:234}}.
		\STATE {Update $\nu_{\phi_n\to \bar{x}_j}$ and $\mu_{\phi_n\to \bar{x}_j}$ by \eqref{eq:226b}, \eqref{eq:27}}.
		\STATE {Update  $\nu_{\bar{x}_j\to \phi_n}$ and $\mu_{\bar{x}_j\to \phi_n}$ by \eqref{eq:210}, \eqref{eq:211}}.
        \STATE {Update  $\nu_{\bar{x}_j\to \mathcal {A}_j}$ and $\mu_{\bar{x}_j\to \mathcal {A}_j}$  by  \eqref{eq:212}, \eqref{eq:213}}.
		\STATE { \textbf{\% Update transform layer based on  FN  $\mathcal{A}$}}			

		\STATE {Update  ${\bm{\nu}}_{\mathcal A\to \mathbf{x}}$ and ${\bm{\mu}}_{\mathcal A\to \mathbf{x}}$ by \eqref{eq:214}.}
		\STATE {Update  ${\bm{\nu}}_{\rm{x}}$ and ${\bm{\mu}}_{\rm{x}}$ by \eqref{eq:2190}.}	        
		\STATE {Update  ${\bm{\nu}}_{\mathcal A\to {\mathbf{\bar x}}}$ and ${\bm{\mu}}_{\mathcal A\to {\mathbf{\bar x}}}$ by \eqref{eq:215}}.

		\STATE {\textbf{\% BEM channel estimation}}				
		\STATE {Update  ${\bm\nu}^{l}_c$ and ${\bm\mu}^{l}_c$ at $R=2$ by \eqref{eq:512}, \eqref{eq:513}.}
		\STATE {Update $\mathbf{\bm{\nu}}^l_{\rm{h}}$ and $\mathbf{\bm{\mu}}^l_{\rm{h}}$ at $R=2$ by  \eqref{eq:516}}.
		
		\IF{
			$\displaystyle \frac{\|{\bm{\mu}}^{l+1}_{\rm{h}} - \mathbf{\bm{\mu}}^{l}_{\rm{h}}\|^2}{\|\mathbf{\bm{\mu}}^{l}_{\rm{h}}\|^2} < \epsilon$ \textbf{and} $\displaystyle \frac{\|\mathbf{\bm{\mu}}^{l+1}_{\rm{x}} - \mathbf{\bm{\mu}}^{l}_{\rm{x}}\|^2}{\|\mathbf{\bm{\mu}}^{l}_{\rm{x}}\|^2} < \epsilon$}
			\STATE {break}
		\ENDIF
		\STATE {$l = l+1$}.
		\ENDWHILE
		\ENSURE  
			$\mathbf{\bm{\mu}}_{\rm{x}}, \mathbf{\bm{\mu}}_{\rm{h}}$. \\
	\end{algorithmic}
\end{algorithm}
\vspace{-0.5cm}
\section{Performance Analysis}
\label{sec:5} 
This section delves into the lower bound on the channel estimation error of the proposed BEM-assisted estimator, the MSE performance  of the proposed MLMP algorithm, and the computational complexity analysis of the proposed BEM-MLMP-JCED scheme.
\vspace{-0.5cm}
\subsection{Lower Bound on Channel Estimation Error}
The theoretical channel estimation   lower bound is achieved when the data vector  $\mathbf{x}_{\rm{d}}$  is correctly decoded and  refined for channel estimation. 

According to \eqref{eq:51}, the DAFT-domain received signal can be  rewritten as
 \begin{equation}
	\scalebox{0.9}{$%
		\begin{aligned}
			\mathbf{r} =\mathbf{Dc}+ \mathbf{\hat e}_{\rm{mod}}+ \mathbf{\hat{w}}
		\end{aligned},
		$}
	\label{eq:5001}   
\end{equation}
where $\mathbf{D} = [\mathbf{d}_{0,0},\dots,\mathbf{d}_{0,L-1},\dots,\mathbf{d}_{Q,L-1}]\in \mathbb{C}^{N \times L(Q+1)} $,
$\mathbf{d}_{q,i} = \mathbf{A}{\rm{diag}}\{\mathbf{u}_q\}\mathbf{G}\mathbf{F}^H{\rm{diag}}\{\mathbf{F}\mathbf{A}^H\mathbf{x}\}\mathbf{F}(:,i)\in \mathbb{C}^{N \times 1}$.

In this paper, we assume a Jakes' Doppler model. Based on \eqref{eq:5001}, we first introduce the following proposition regarding  the modeling error  $\hat{\mathbf{e}}_{\rm mod}$. 

\textit{Proposition 1:}
  Denote $\mathbf{h}_\ell$ by the channel coefficient
vector corresponding to the $\ell$th path, which is modeled as $\mathbf{h}_\ell = \sqrt{\bar P_\ell}\,\tilde{\mathbf{h}}_\ell$ where $\bar P_\ell$ is the average power of the $\ell$th path and 
$\tilde{\mathbf{h}}_\ell$ is a zero-mean, unit-variance complex Gaussian vector. All paths share the same normalized Jakes' covariance
matrix $\mathbf{R}_{\rm h} \triangleq \mathbb{E}[\tilde{\mathbf{h}}_\ell
\tilde{\mathbf{h}}_\ell^{H}]$.
Then  $\hat{\mathbf{e}}_{\rm mod}$ is also a
zero-mean complex Gaussian vector, and its variance is given by
\begin{equation}
	\scalebox{0.9}{$%
		\begin{aligned}
			\sigma_{\rm{e}}^2 = \frac{1}{N}{\rm{Tr}}\left(\mathbf{ {U}}_{\rm{I}}\mathbf{R}_{\rm{h}}\mathbf{ {U}}_{\rm{I}}^H\right) \sum_{\ell=0}^{L-1}\bar P_\ell\sum_{k=0}^{N-1} |\lambda_\ell[k]|^{2}|z_k|^{2},
		\end{aligned}
		$}
	\label{eq:702}   
\end{equation}
where $\mathbf{ {U}}_{\rm{I}} = \left(\mathbf I_{N}-\mathbf{U}(\mathbf {U}^{H}\mathbf {U})^{-1}\mathbf { U}^{H}\right)$, $z_k$ is the $k$th element of
$\mathbf{z} = \mathbf{F}\mathbf{A}^{H}\mathbf{x}$,  and $\lambda_\ell[k]$ denotes the $k$th eigenvalue of the matrix $\mathbf{G}_\ell = \mathbf{G}\mathbf{\Pi}^\ell$.

\textit{Proof}: See Appendix \ref{appe1}. \hfill $\blacksquare$ 

With  \eqref{eq:702}, the Fisher information matrix  of $\mathbf{c}$ can be expressed as \cite{pakroohAnalysisFisherInformation2015}
\begin{equation}
	\scalebox{0.9}{$%
		\begin{aligned}
			\mathbf{J(c)} = (\sigma_{\rm{e}}^2+\sigma^2)^{-1}\mathbf{D}^H  \mathbf{D}.
		\end{aligned}
		$}
	\label{eq:7}   
\end{equation}
Accordingly, the  covariance of any unbiased estimator of $\mathbf{c}$ is given by
\begin{equation}
	\scalebox{0.9}{$%
		\begin{aligned}
			{\rm{Cov}}\mathbf{(\hat{c})} \succeq  (\sigma_{\rm{e}}^2+\sigma^2)(\mathbf{D}^H  \mathbf{D})^{-1}.
		\end{aligned}
		$}
	\label{eq:71}   
\end{equation}
Based on  \eqref{eq:71} and  \eqref{eq:802}, the covariance matrix of any unbiased estimator  of $\mathbf h$ satisfies
\begin{equation}
	\scalebox{0.9}{$%
		\begin{aligned}
			{\rm{Cov}}\mathbf{(\hat{h})} \succeq \mathbf{\tilde U}{\rm{Cov}}\mathbf{(\hat{c})}\mathbf{\tilde U}^H.
		\end{aligned}
		$}
	\label{eq:8023}   
\end{equation}
Furthermore, one has the covariance of $\mathbf{{e}}_{\rm{mod}}$  given by
\begin{equation}
	\scalebox{0.9}{$%
		\begin{aligned}
			\mathbf{R}_{{\rm{e}}}= \mathbb{E}\left[ \mathbf{{e}}_{\rm{mod}}\mathbf{{e}}_{\rm{mod}}^H\right]=\mathbf{\tilde {U}}_{\rm{I}}\mathbf{\hat{R}}_{\rm{h}}\mathbf{\tilde {U}}_{\rm{I}}^H
			=\left(\mathbf{ {U}}_{\rm{I}}\mathbf{R}_{\rm{h}}\mathbf{ {U}}_{\rm{I}}^H\right)\otimes \mathbf{P}_{\rm{h}},
		\end{aligned}
		$}
	\label{eq:806}   
\end{equation}
where $\mathbf{\hat{R}}_{\rm{h}}= \mathbf{R}_{\rm{h}}\otimes \mathbf{P}_{\rm{h}}$ is the correlation matrix of $\mathbf{h}$, $\mathbf{P}_{\rm{h}} = {\rm{diag}}\{\bar P_0,\bar P_1,\cdots,\bar P_{L-1}\}$ represents a diagonal matrix consisting of the power $\bar P_\ell$ of the $\ell$th path.

By applying  \eqref{eq:8023} and \eqref{eq:806}, the lower bound on the Normalized MSE (NMSE) of the channel estimation can be expressed as 
\begin{equation}
	\scalebox{0.9}{$%
		\begin{aligned}
			{\rm{NMSE}_{LB}} =  \frac{(\sigma_{\rm{e}}^2+\sigma^2) {\rm{Tr}}\left(\mathbf{\tilde U}(\mathbf{D}^H  \mathbf{D})^{-1}\mathbf{\tilde U}^H\right) + {\rm{Tr}}(\mathbf{R}_{{\rm{e}}})}{{\rm{Tr}}(\mathbf{R}_{\rm{h}})}.
		\end{aligned}
		$}
	\label{eq:701}   
\end{equation}
\vspace{-0.5cm}
\subsection{State Evolution}
In this subsection, we analyze the MSE performance of the proposed algorithm based on state evolution. 
Owing to the structural similarities across different layers, their analyses are consolidated in a unified framework.

According to \eqref{eq:061}, the received TD signal  can be rewritten as
 \begin{equation}
	\scalebox{0.9}{$%
		\begin{aligned}
			\mathbf{y}  = \mathbf{H}_{\rm{all}}\mathbf{x} + \mathbf{w},
		\end{aligned}
		$}
	\label{eq72}   
\end{equation}
where $\mathbf{H}_{\rm{all}} = \mathbf{HGA}^H$ denotes the overall equivalent TD channel matrix.

Let $ \hat{\mathbf x}^{l}$ be the estimate of $\mathbf x$ at the $l$th iteration. We define the estimation error as $\mathbf{e} = \mathbf{x}-\hat{ \mathbf{x}}^{l}$, 
and the corresponding MSE at the $l$th iteration is given by
\begin{equation}
	\scalebox{0.9}{$%
		\begin{aligned}
			\eta_x^{l}  = \frac{1}{N} \mathbb{E}\left[ \Vert \mathbf x- \hat{\mathbf x}^{l} \Vert^2\right] = \frac{1}{N} \mathbb{E}\left[ \Vert\mathbf{e} \Vert^2\right].
		\end{aligned}
		$}
	\label{eq73}   
\end{equation}
The residual passed to the denoiser can be modeled as an equivalent scalar AWGN channel. Specifically, there exists a linear projection that yields a scalar observation $\hat x$ with effective noise variance $ \bar{\nu}_x^{l} $ such that
 \begin{equation}
	\scalebox{0.9}{$%
		\begin{aligned}
			\hat x = x + \sqrt{\bar{\nu}^{l}_x}z_x, \quad
			z_x\sim \mathcal{CN}\left(0, 1 \right).
		\end{aligned}
		$}
	\label{eq79}   
\end{equation}

\textit{Proposition 2:} {Under the standard state evolution model, where the input to the denoiser is obtained from a back-projected residual, $\bar{\nu}_x^{l}$ is given by
}
 \begin{equation}
	\scalebox{0.9}{$%
		\begin{aligned}
			\bar{\nu}^{l}_x = \frac{\eta_x^{l}}{N} {\rm{Tr}}\Bigg[\Big(\mathbf{I}_N-\frac{1}{\tau_{\rm{H_{all}}}}\mathbf{H}_{\rm{all}}^H\mathbf{H}_{\rm{all}}\Big)^2\Bigg] + \frac{\sigma^2}{\tau_{\rm{H_{all}}}},
		\end{aligned}
		$}
	\label{eq80}   
\end{equation}
where $\tau_{\rm{H_{all}}} = \frac{1}{N}{\rm{Tr}}(\mathbf{H}_{\rm{all}}^H\mathbf{H}_{\rm{all}})$. 

\textit{Proof}: See Appendix \ref{appe2}. \hfill $\blacksquare$ 

For the equivalent scalar channel in \eqref{eq79}, the denoiser implements the MMSE estimator of $x$ given $\hat x$. The corresponding posterior MSE depends only on  $\bar{\nu}_x^{l}$ and is characterized by the   posterior   MMSE function 
 \begin{equation}
	\scalebox{0.9}{$%
		\begin{aligned}
        \eta_{\mathrm{post}}^{l}=\mathrm{mmse}\left(\bar{\nu}^{l}_x\right)
			=\mathbb{E}\left[\left|x-\mathbb{E}(x|\hat x)\right|^2\right].
		\end{aligned}
		$}
	\label{eq81}   
\end{equation}
It should be noted that \eqref{eq81} is a nonlinear function of the  signaling constellation  $\mathbb M$, and it is  challenging to obtain the closed-form expression. Consequently, the average posterior variance is approximated using  Monte Carlo simulation.

Finally, based on the damping factor $\gamma$, the updated $\eta_x^{l+1}$ is given as
 \begin{equation}
	\scalebox{0.9}{$%
		\begin{aligned}
			\eta_x^{l+1}
			&= (1-\gamma) \eta_x^{l} + \gamma \eta_{\mathrm{post}}^{l}.
		\end{aligned}
		$}
\label{eq82}   
\end{equation}
\vspace{-0.5cm}
\subsection{Complexity Analysis}
The computational complexity of the TD channel and FTN layers are $\mathcal{O}(NL)$ and $\mathcal{O}(N(2N_l+1))$, respectively. 
The transform layer requires $\mathcal{O}(N\log N)$ operations, since the multiplication by  $\mathbf{A}$ is implemented via the fast Fourier transform.
The complexity of BEM-assisted channel estimation can be divided into two parts. Specifically, 
the update in \eqref{eq:514} has a complexity of $\mathcal{O}(N(L(Q+1))^2)$, while the matrix inversion in \eqref{eq:512}–\eqref{eq:513} has a complexity $\mathcal{O}((L(Q+1))^3)$, which is negligible compared with the $\mathcal{O}(N)$ term for large $N$.
In addition, for an $M$-ary modulation alphabet, the operations in \eqref{eq:216}–\eqref{eq:218} have a complexity of $\mathcal{O}(NM)$.
 Hence,  the overall computational complexity of Algorithm \ref{alg1}  can be approximated as
\begin{equation}
	\scalebox{0.9}{$%
		\begin{aligned}
		\mathcal{O}\left(I_tN\left(({L}(Q+1))^2+\log N+2N_l\right)\right)
		\end{aligned}
		$},
	\label{eq:8}   
\end{equation}
where $I_t$ denotes the number of iterations. 
The proposed algorithm exhibits computational complexity that scales linearly with $N$, offering a substantial improvement over the conventional LMMSE approach, whose complexity is on the order of $\mathcal{O}(N^3)$.
\vspace{-0.5em}
\section{Simulation Results}
\label{sec:6} 
This section conducts comprehensive  simulations   to validate the performance of  the proposed AFDM-FTN system.   We consider $L$ independent paths, where each path coefficient is modeled as a zero-mean complex Gaussian random variable with variance  of $1/L$. The normalized Doppler shift for the $i$th path is generated as $\hat{f}_i = \hat{f}_{\max}\cos(\theta)$, where  $\theta$ follows a uniform distribution in $[-\pi,\pi]$.
Both $h_{\rm{tx}}(t)$ and $h_{\rm{rx}}(t)$ are root raised cosine (RRC) filters. 
\textcolor{blue}{Unless otherwise specified, the simulation parameters summarized in Table I are considered, where $L$ is determined by the adopted channel profile and resolvable multipath components, without depending on the FTN compression factor $\alpha$. The BEM order $Q$ is selected according to the $\hat f_{\max}$, $N$, and $R$. In this setting, $Q=6$ is sufficient to model the time-varying channel \cite{liuNearOptimalBEMOTFS2022}, while increasing $Q$ only enlarges the estimation dimension $L(Q+1)$ with limited performance gain. Therefore, $Q$ is governed by the channel time variation, whereas $\alpha$ mainly controls the FTN-ISI strength and nominal SE gain.}  
\begin{table}[t]
	\caption{Simulation parameters}
	\label{table_1}
	\centering
	\renewcommand{\arraystretch}{1.2}
	\begin{tabular}{c|c}
		\hline                     
		\textbf{Parameters} & \textbf{Values} \\ \hline \hline
		Carrier frequency (GHz)              & 24 \\ \hline 
		Number of paths $L$                     & 3 \\ \hline
		Relative movement speed (km/h)       & 675 \\ \hline
		AFDM subcarrier spacing $\Delta f$ (kHz)        & 15 \\ \hline
		Number of AFDM subcarriers $N$       & 1024 \\ \hline
		FTN compression factor $\alpha$                  & \textcolor{blue}{1, 0.9, 0.8, 0.769, 0.703} \\ \hline
		FTN-ISI effective length ${N}_l$     & 5 \\ \hline
		RRC roll-off factor  $\beta$        & 0.3 \\ \hline
		Pilot-to-noise ratio ${\rm{SNR}}_p$ (dB) & 20 \\ \hline
		Maximum Doppler shift $f_{\rm{max}}$  (kHz)          & 15 \\ \hline
		BEM order $Q$                        & 6 \\ \hline
		Modulation scheme                    & \textcolor{blue}{QPSK and 16-QAM} \\ \hline
	\end{tabular}
\end{table}
\vspace{-0.5em}
\subsection{Perfect Channel Information}
\begin{figure}[t]
	\centering
	\includegraphics[width=0.85\columnwidth]{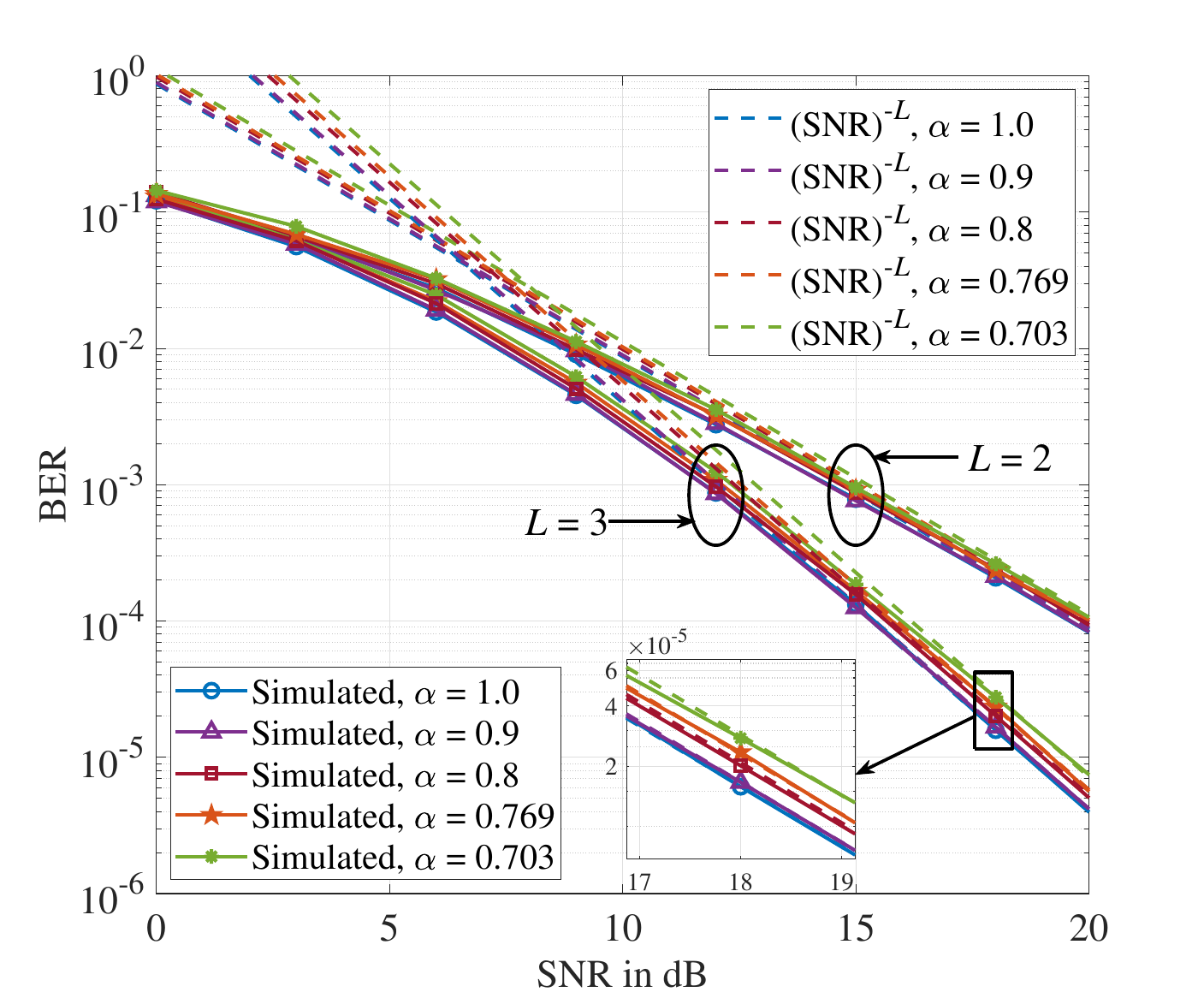}
	\caption{\textcolor{blue}{BER performance based on ML detection under perfect channel.}}
	\label{fig5}
    \vspace{-0.5cm}
\end{figure}

We first present the BER performance of the AFDM-FTN under perfect channel information, as shown in Fig. \ref{fig5}, where $N=16$, $L=2,3$, and a maximum likelihood detector are used. It should be noted that, for \(\alpha=1\), the AFDM-FTN corresponds to the conventional AFDM system. 
\textcolor{blue}{The Mazo limit indicates that FTN signaling can reduce the symbol interval while maintaining the same minimum Euclidean distance as the Nyquist signaling system under ideal detection conditions. For the adopted RRC pulse with $\beta =0.3$, the Mazo-limit-related value for single-carrier FTN signaling is $\alpha =0.703$ \cite{liverisExploitingFasterthannyquistSignaling2003}. Hence, this paper considers $\alpha =0.703$ as a special case to examine the system performance near the Mazo limit boundary.}
The dashed lines are the   asymptotic BER curves based on the pairwise error probability (PEP) based analysis, whose   slopes are $  (\text{SNR})^{-L} $ \cite{luoAFDMSCMAPromisingWaveform2024}, \cite{liAffineFrequencyDivision2025a}, and \cite{9369968}.
The main observations of Fig. \ref{fig5} are summarized as follows: 1) At high SNRs, the simulated BER curves closely match the analytical asymptotic results, indicating that the proposed AFDM-FTN scheme can fully exploit the available channel diversity; 2) The BER performance improves as the number of propagation paths increases, owing to the additional multipath diversity gain; \textcolor{blue}{3) as $\alpha$ decreases, BER degrades as FTN-ISI becomes more severe. Nevertheless, the degradation remains moderate in the considered compression range. In particular, even for $\alpha=0.703$, the performance loss with respect to $\alpha=1.0$ remains within approximately $1$ dB at $\text{BER} = 2\times 10^{-5}$. Therefore, the proposed AFDM-FTN can still achieve comparable BER in the considered compression region, while providing additional SE gain. Based on the above observations, $\alpha=0.8$ is adopted in the subsequent simulations as a representative FTN compression factor, since it provides a relatively aggressive SE improvement while maintaining BER performance close to the Nyquist case.}
\vspace{-0.5em}
\subsection{The Proposed BEM-MLMP-JCED}
Next, we present simulation results of the proposed  BEM-MLMP-JCED.
We consider $L=3$ and assume each path $l_i$ is random generated from   $[0, 5]$. 
The damping factor is set as $\gamma = 0.6$.

\begin{figure}[t]
	\centering		
	\subfigure[\textcolor{blue}{QPSK.}]{
		\label{QPSK}
		\includegraphics[width=0.48\columnwidth]{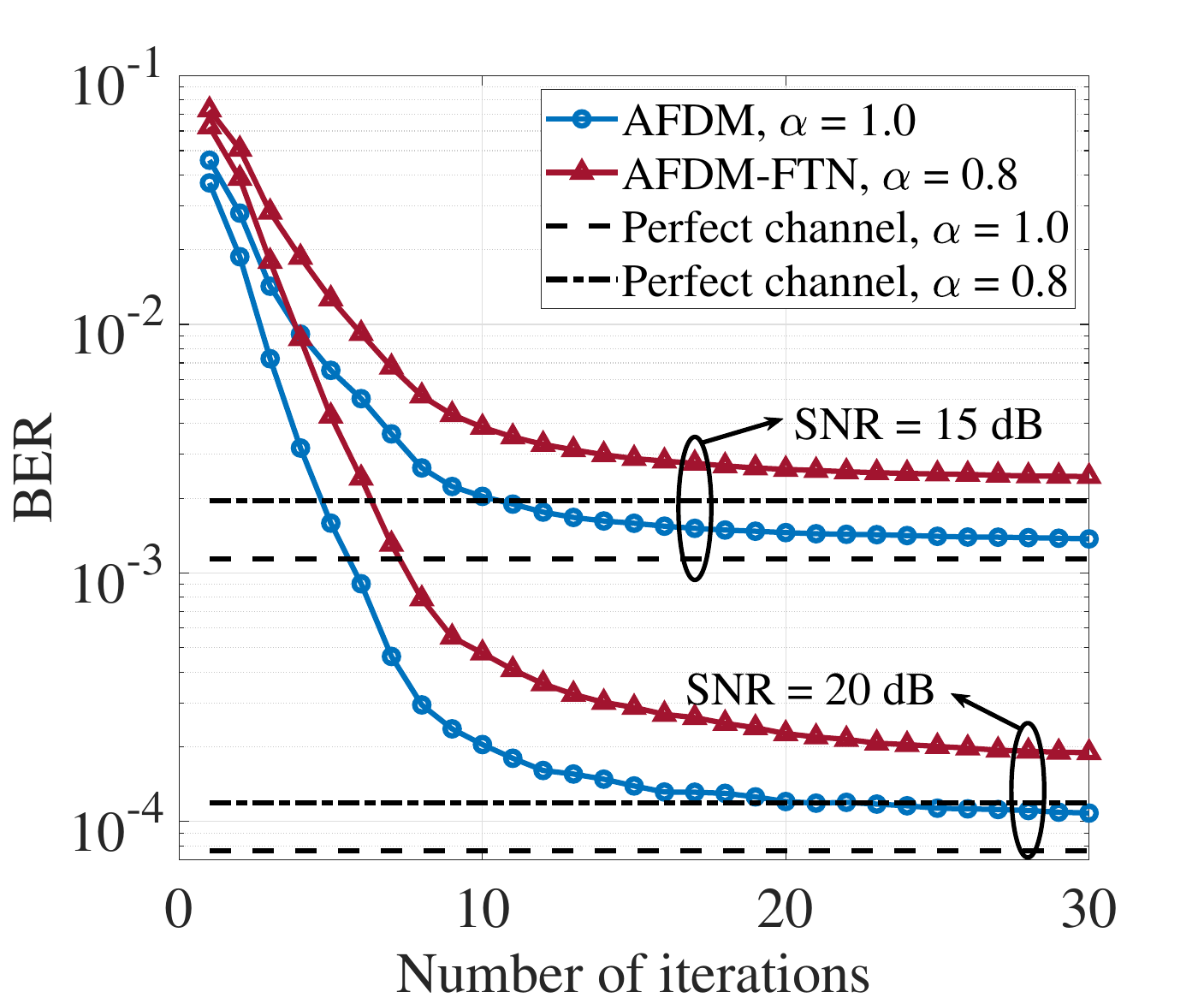}}	
	\subfigure[\textcolor{blue}{16-QAM.}]{
		\label{16QAM}
		\includegraphics[width=0.48\columnwidth]{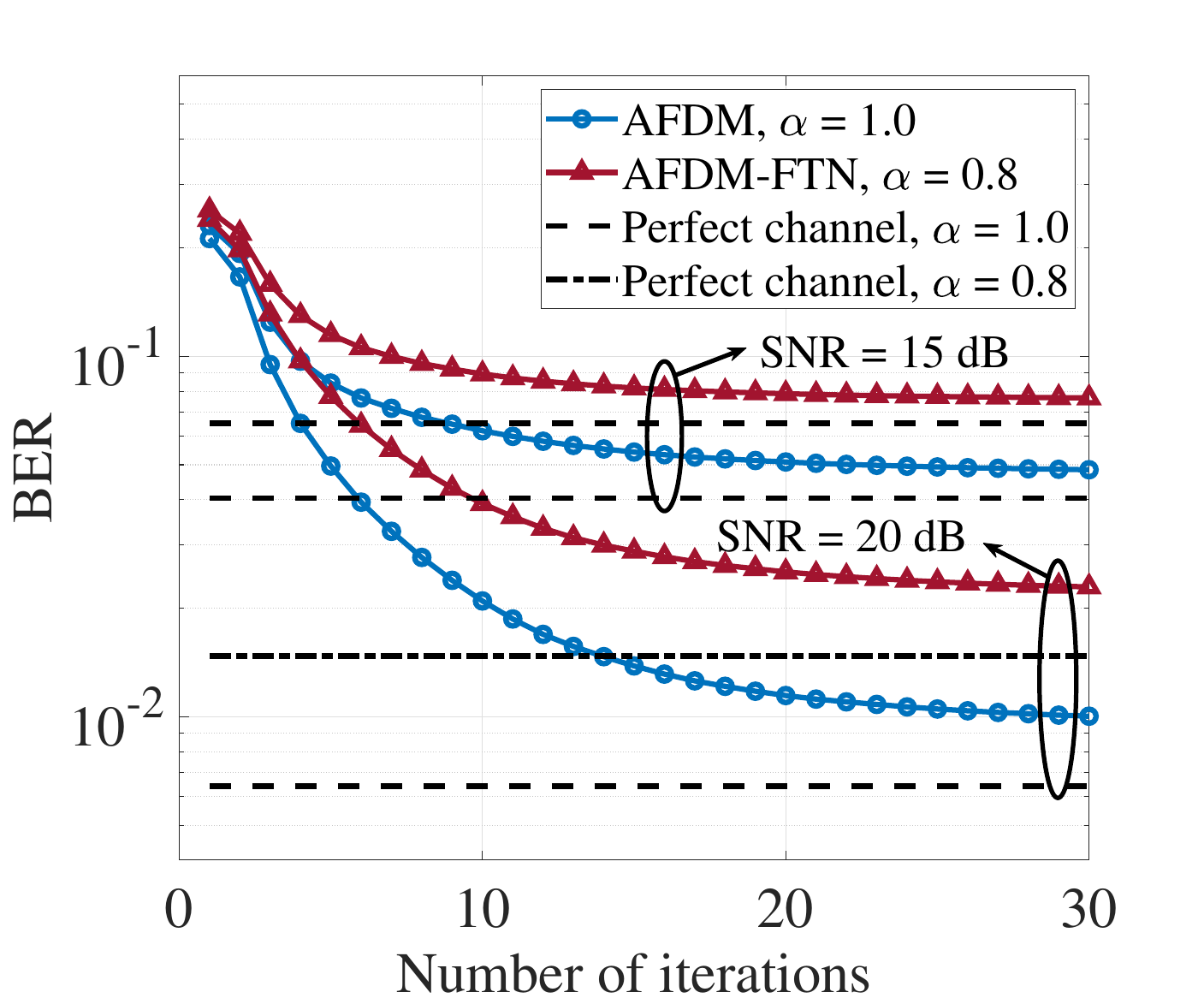}}		
	\caption{\textcolor{blue}{BER convergence of the proposed BEM-MLMP-JCED algorithm.}}
	\label{fig8}
	\vspace{-0.5cm}
\end{figure}

\textcolor{blue}{Fig. \ref{fig8} shows the   convergence behaviors of the proposed BEM-MLMP-JCED algorithm for QPSK and 16-QAM at $\text{SNR} = 15$ dB and $\text{SNR} = 20$ dB}. We set  the   maximum number of iterations and the stopping threshold as  $I_{\rm{max}} = 30$ and $\epsilon = 10^{-5}$, respectively. 
Overall, the proposed AFDM-FTN system with the BEM-MLMP-JCED algorithm converges within approximately $I_t = 20$ iterations. More importantly, the proposed BEM-MLMP-JCED scheme achieves performance close to that obtained with perfect channel state information. This indicates that the improved data detection in each iteration provides increasingly reliable soft information for channel estimation, which in turn enhances subsequent detection, thereby improving the overall system performance.
\textcolor{blue}{Compared with QPSK, 16-QAM has a smaller minimum Euclidean distance between constellation points and is therefore more sensitive to residual channel estimation error, FTN-ISI, and BEM modeling error. As shown in Fig. \ref{16QAM}, although 16-QAM exhibits  a marginally larger convergence gap, the proposed BEM-MLMP-JCED algorithm still exhibits stable iterative convergence and approaches the benchmark with perfect channel. This confirms that the proposed iterative channel estimation and data detection mechanism remains effective for higher order modulation.} 
\textcolor{blue}{It is also observed that the gap between the proposed algorithm and the perfect channel benchmark becomes more noticeable at high SNR. This is because the dominant impairment changes from additive noise to residual modeling and estimation errors, including BEM approximation error, finite FTN-ISI truncation error, and imperfect data aided channel refinement. As the noise level decreases, these residual errors become the main source of the remaining performance gap.}

\begin{figure}[t]
	\centering
	\includegraphics[width=0.85\columnwidth]{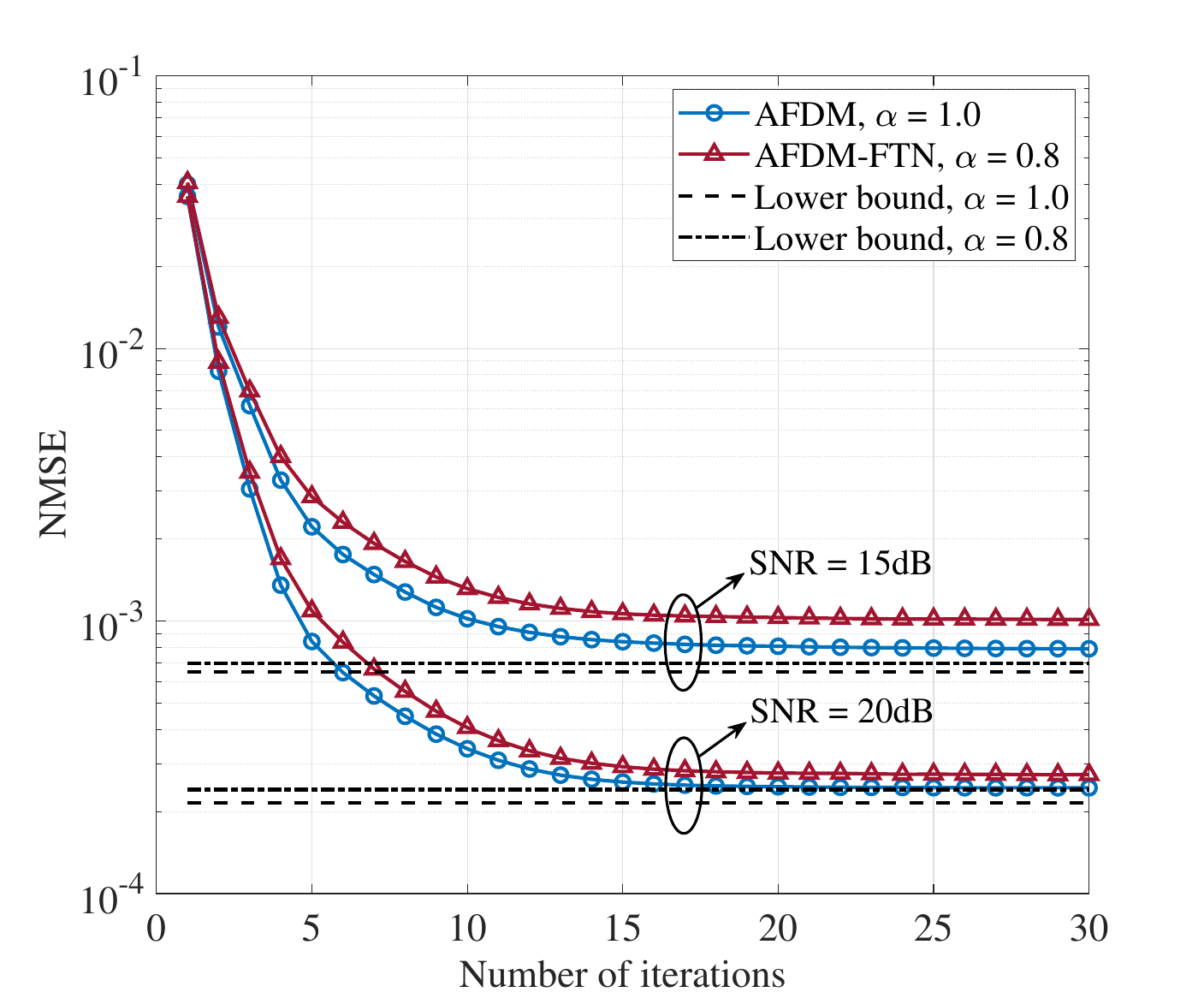}
	\caption{NMSE convergence of the proposed BEM-MLMP-JCED algorithm.}
	\label{fig9}
    \vspace{-0.2cm}
\end{figure}

\textcolor{blue}{Fig. \ref{fig9} further illustrates the NMSE convergence behavior of the proposed BEM-MLMP-JCED algorithm, where QPSK is used as the representative modulation setting to focus on the channel estimation convergence and its agreement with the derived lower bound.}
The initial channel estimate is obtained using a BEM order of $R = 1$, after which a refined estimate is computed by increasing the order to $R = 2$. As observed, the NMSE decreases sharply during the early iterations and exhibits a convergence trend similar to that of the BER performance. Eventually, the NMSE converges to the theoretical lower bound given in \eqref{eq:701}. Moreover, the gap between the simulated NMSE and the derived lower bound becomes smaller at higher SNR values, e.g., $\text{SNR} = 20$ dB. 

\begin{figure}[t]
	\centering
	\includegraphics[width=0.85\columnwidth]{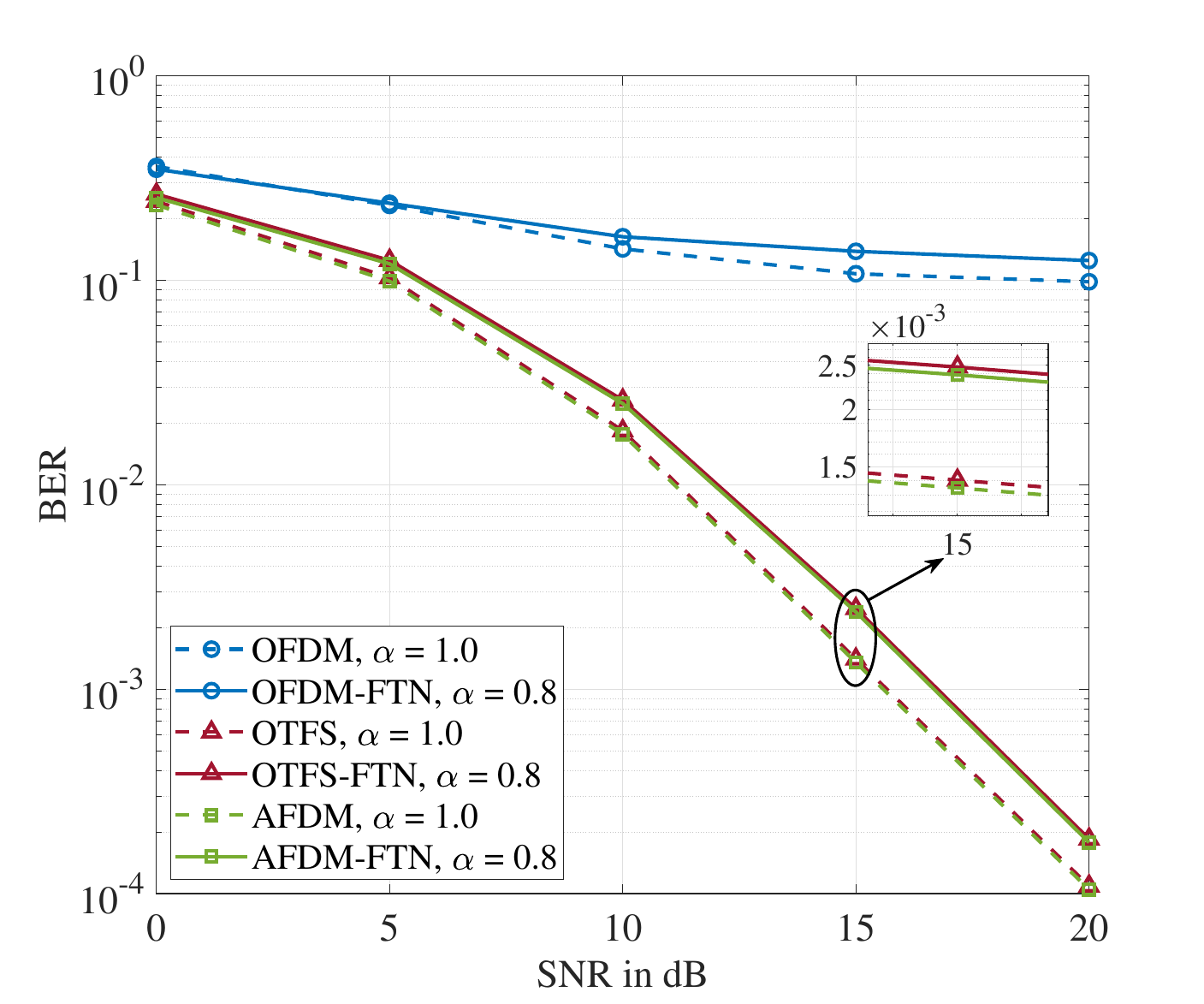}
	\caption{\textcolor{blue}{Comparison of BER performance under different waveforms.}}
	\label{fig6}
    \vspace{-0.5cm}
\end{figure}

\textcolor{blue}{Fig. \ref{fig6} compares the BER performance of the proposed AFDM-FTN, OFDM-FTN, and OTFS-FTN with the proposed BEM-MLMP-JCED algorithm, where the OTFS-FTN scheme in \cite{tongAdaptiveFTNSignaling2025} is included as a comparison scheme for doubly selective channels.
For a fair comparison, we consider an OTFS frame consisting of $N_{\rm OTFS}$ time bins and $M_{\rm OTFS}$ frequency samples, where the sampling intervals $T_{\rm OTFS}$ and $\Delta f_{\rm OTFS}$ satisfy $T_{\rm OTFS}\Delta f_{\rm OTFS}=1$. The AFDM block length is set as $N=N_{\rm OTFS}M_{\rm OTFS}$, so that one AFDM block and one OTFS frame contain the same number of transmitted symbols, including data, pilot, and guard symbols. In the simulations, $N=1024$ and $N_{\rm OTFS}=M_{\rm OTFS}=32$ are adopted.
For FTN-based schemes, the same compression factor, pulse shaping filter, and roll-off factor are used, so that the nominal SE gain is identical.}
It can be observed that  the proposed  AFDM-FTN achieves slightly better BER performance than OTFS-FTN  with the proposed BEM-MLMP-JCED.  Both  AFDM-FTN and OTFS-FTN   demonstrate  a notable performance improvement compared with the OFDM-FTN scheme. This performance improvement arises mainly from two factors. First, AFDM can effectively exploit additional diversity in doubly selective channels. Second, under FTN signaling, OFDM suffers from frequency-response nulls that lead to irreversible information loss on certain subcarriers, while AFDM interprets the FTN effect as a form of multipath superposition, thereby enabling more reliable data detection.

\begin{figure}[t]
	\centering
    \includegraphics[width=0.85\columnwidth]{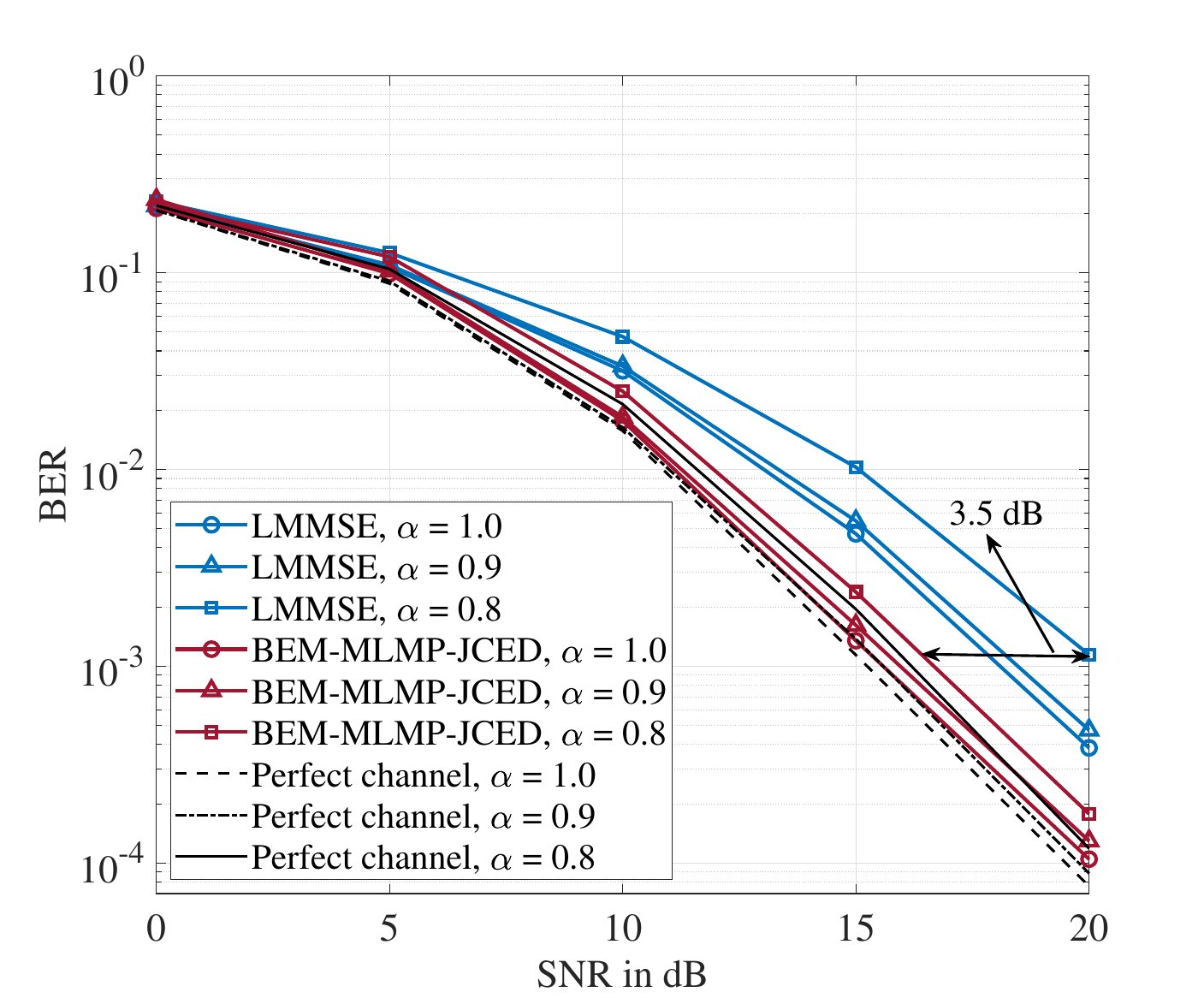}
	\caption{Comparison of BER performance under different detection algorithms.}
	\label{fig7}
    \vspace{-0.5cm}
\end{figure}

Fig. \ref{fig7} compares the BER performance of the proposed BEM-MLMP-JCED algorithm with that of the LMMSE detector for uncoded AFDM-FTN systems. 
For reference, the curve of the MLMP detector with perfect channel is also shown as a benchmark.
First, it is interesting to observe that the proposed BEM-MLMP-JCED with $\alpha = 0.8$ achieves BER performance comparable to that of $\alpha = 1$, indicating strong robustness to bandwidth compression. 
In contrast, the LMMSE detector suffers from pronounced performance degradation, particularly as $\alpha$ decreases. 
Compared with the perfect channel benchmark, the proposed channel estimation scheme attains almost the same BER performance, demonstrating the effectiveness of the proposed BEM-assisted estimator.
Furthermore, the proposed BEM-MLMP-JCED achieves an approximately $3.5$~dB gain over the LMMSE approach at a BER of $1.1 \times 10^{-3}$ and  $\alpha = 0.8$. More importantly, the proposed BEM-MLMP-JCED incurs significantly lower computational complexity than the LMMSE detector, thereby demonstrating comprehensive advantages in both performance and complexity.

\begin{figure}[t]
	\centering
	\includegraphics[width=0.85\columnwidth]{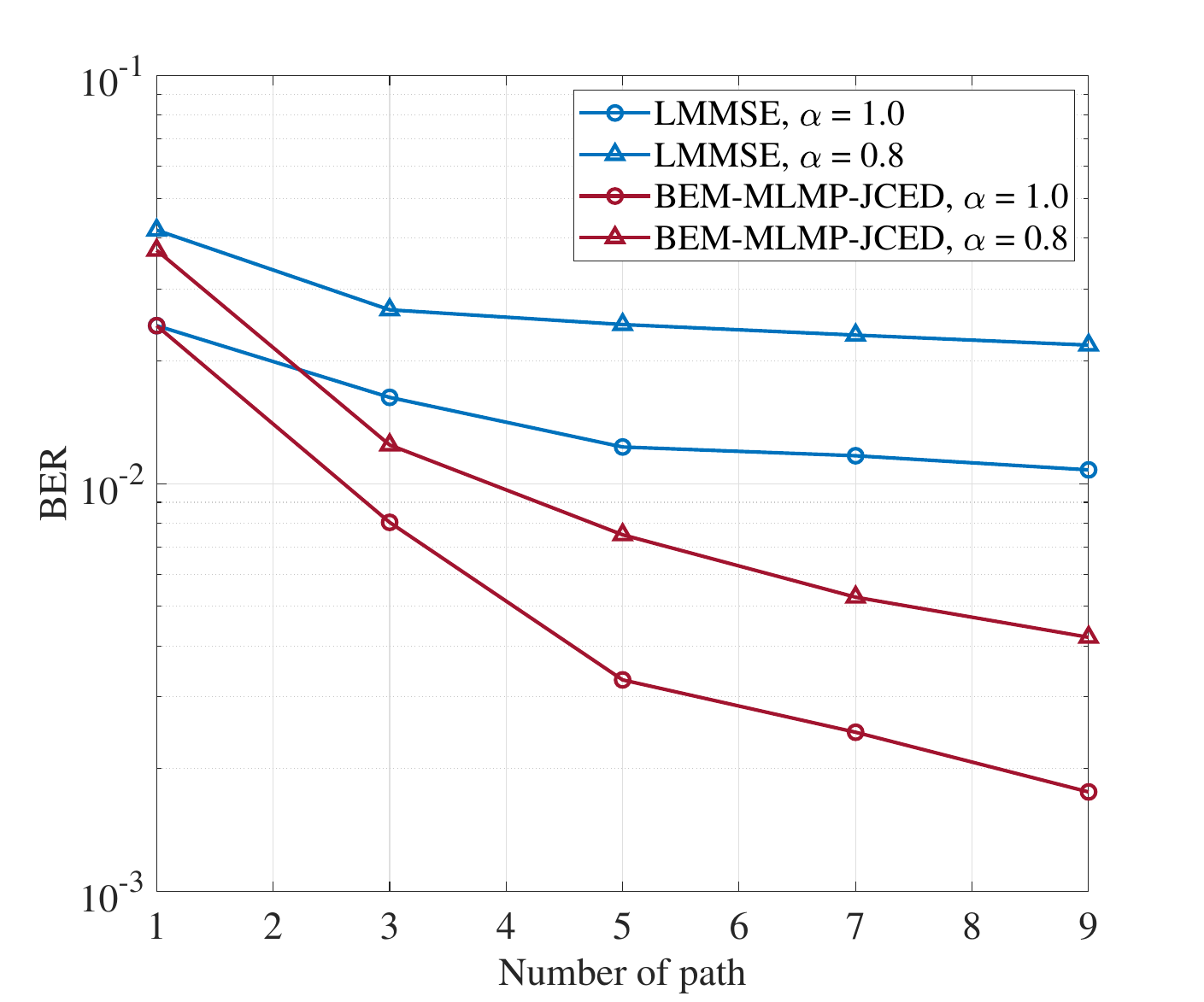}
	\caption{\textcolor{blue}{BER performance versus the number of paths in the AFDM-FTN system at SNR = 12 dB.}}
	\label{fig71}
	\vspace{-0.5cm}
\end{figure}

\textcolor{blue}{Fig. \ref{fig71} further evaluates the BER performance of the proposed algorithm as the number of paths increases under a high-mobility channel with severe Doppler effect. Specifically, the maximum Doppler shift is set to $\hat f_{\max}=1$, and the SNR is fixed at $12$ dB. Both LMMSE and BEM-MLMP-JCED receivers are considered, with equal power assigned to each path and the CPP length chosen to cover the maximum integer delay. It is observed that the BER performance of both receivers improves as the number of paths increases. However, the performance gain of the LMMSE receiver is relatively limited. In contrast, the proposed BEM-MLMP-JCED algorithm achieves a more pronounced BER reduction for both $\alpha=1.0$ and $\alpha=0.8$, owing to its effective exploitation of multipath diversity gain.}

\begin{figure}[t]
	\centering		
	\subfigure[\textcolor{blue}{QPSK.}]{
		\label{QPSKS}
		\includegraphics[width=0.48\columnwidth]{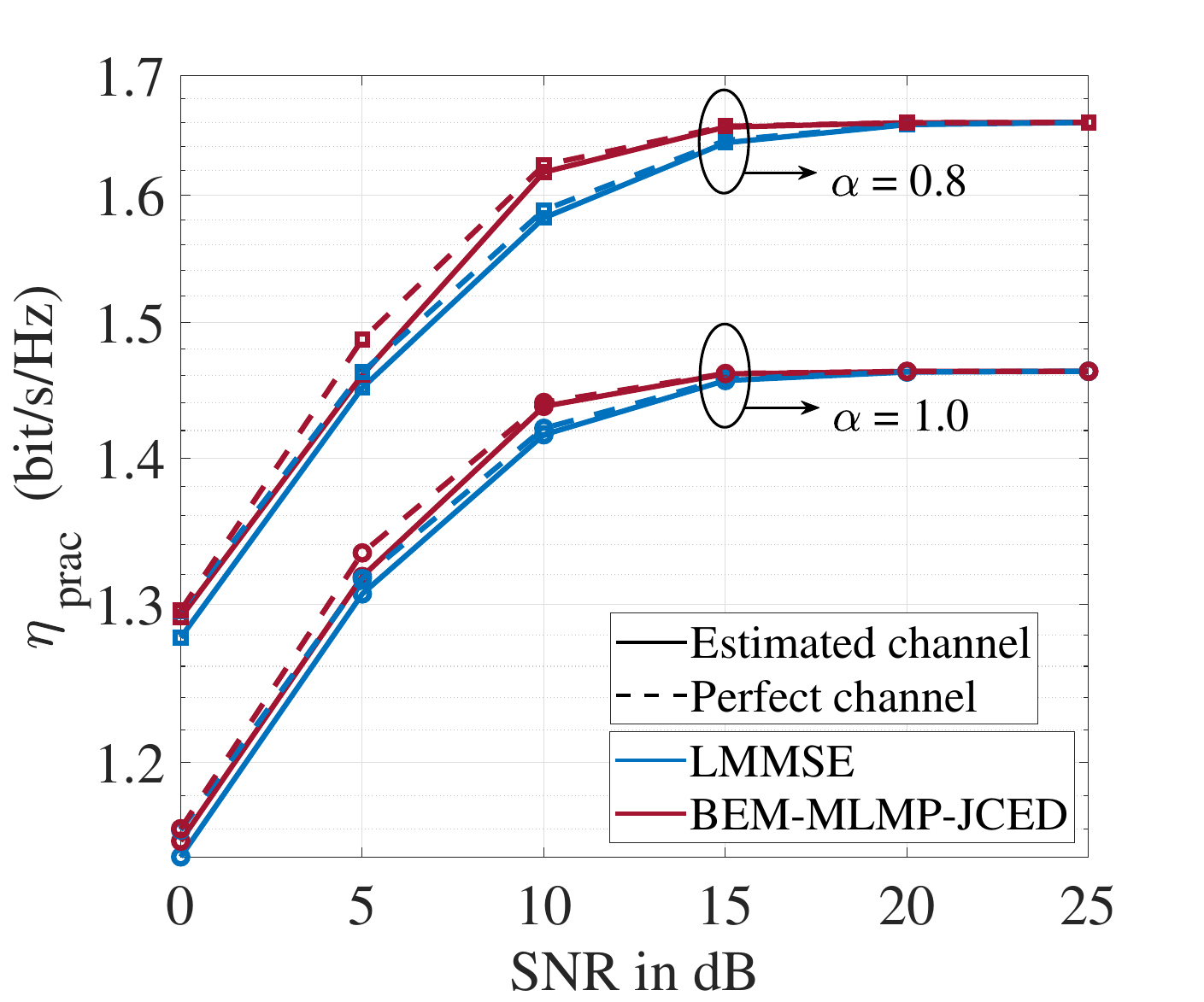}}	
	\subfigure[\textcolor{blue}{16-QAM.}]{
		\label{16QAMS}
		\includegraphics[width=0.48\columnwidth]{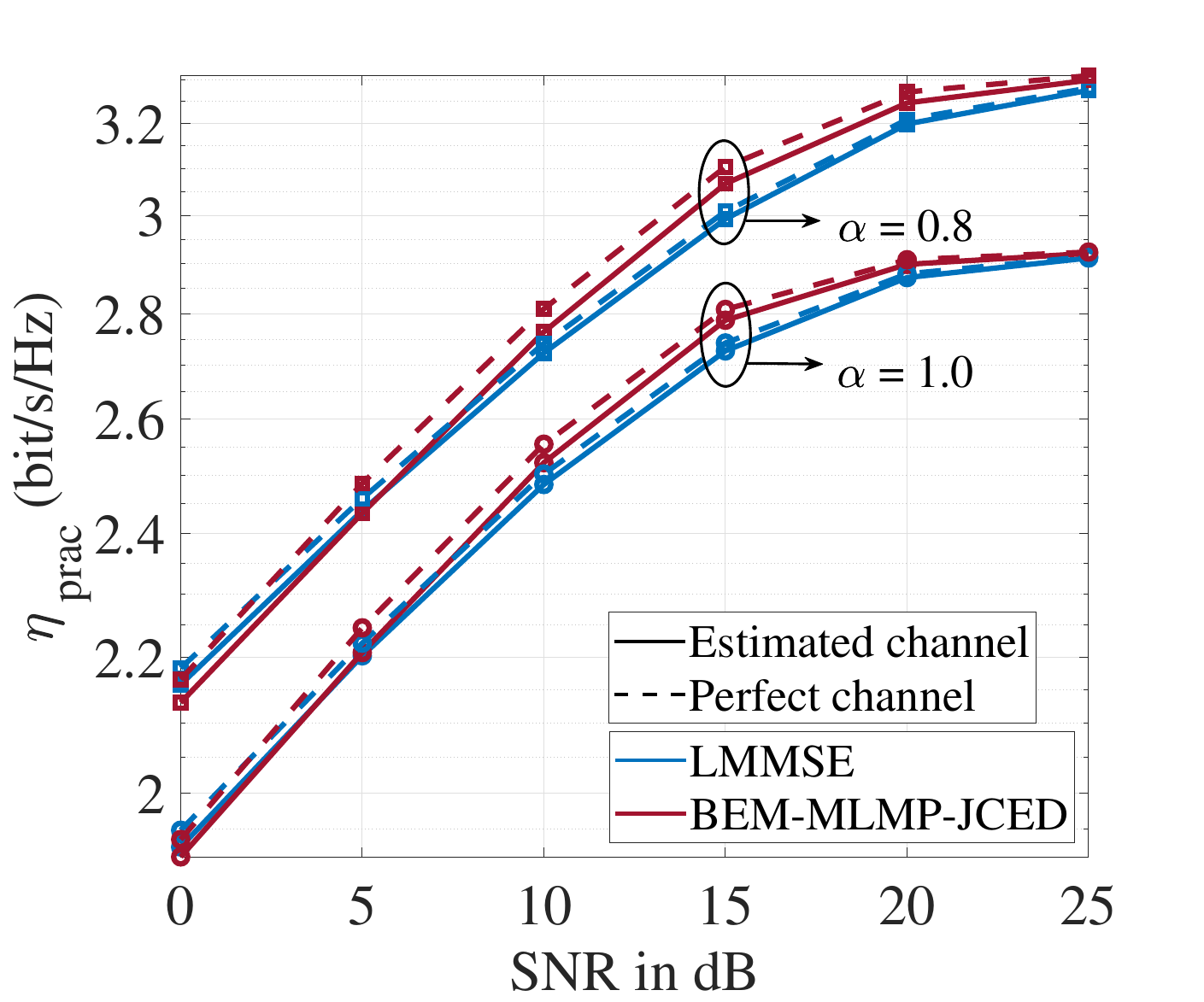}}		
	\caption{\textcolor{blue}{Practical effective SE performance of the proposed  AFDM-FTN system.}}
	\label{fig101}
	\vspace{-0.5cm}
\end{figure}

\textcolor{blue}{In the following, we evaluate the advantages of the proposed AFDM-FTN in terms of practical effective SE. To account for both detection errors and the resources consumed by pilot transmission and guard regions, the practical effective SE is defined as \cite{luoDesignVariableModulation2025}}
\begin{equation}
	\scalebox{0.9}{$%
		\textcolor{blue}{\begin{aligned}
			\eta_{\rm prac}
			=
			(1-\rho_{\rm p,g})
			\frac{r_{\text{c}}\log_2M}{\alpha(1+\beta)}
			(1-\mathrm{BER}),
		\end{aligned}}
		$}
	\label{eq:eta_prac}   
\end{equation}
\textcolor{blue}{where $r_c$ denotes the coding rate, $M$ is the modulation order, $\alpha$ is the FTN compression factor, and $\beta$ is the roll-off factor. The term $\frac{r_c\log_2 M}{\alpha(1+\beta)}$ represents the nominal SE determined by the waveform parameters, $(1-\text{BER})$ accounts for the loss caused by erroneous decoded bits, and $\rho_{\rm p,g}$ denotes the pilot and guard overhead ratio. For the proposed AFDM-FTN scheme, $\rho_{\rm p,g}$ is given by $\rho_{\rm AFDM\mbox{-}FTN}$ defined in Section \ref{sec:3-2}. Hence, $\eta_{\rm prac}$ is used as a link-level effective throughput metric to characterize the practical BER-SE tradeoff. 
Achievable rate and mutual information provide information theoretic characterizations of SE, while \eqref{eq:eta_prac} is adopted here to directly reflect the practical throughput under the considered receiver implementation.
Fig. \ref{fig101} presents the practical effective SE of the proposed uncoded AFDM-FTN system with $r_c=1$. Fig. \ref{QPSKS} shows the QPSK results, while Fig. \ref{16QAMS} gives the corresponding results for 16-QAM. After accounting for pilot and guard resources, the proposed AFDM-FTN retains a clear practical SE advantage over conventional AFDM.
In particular, for QPSK at an SNR of $20$ dB, an additional $0.20$ bit/s/Hz gain is observed for the proposed AFDM-FTN with $\alpha =0.8$ compared with AFDM with $\alpha =1.0$.
For 16-QAM, although the receiver becomes more sensitive to residual channel estimation error and FTN-ISI, the practical SE gain of AFDM-FTN remains evident. In particular, at an SNR of $25$ dB, AFDM-FTN with $\alpha=0.8$ achieves an additional $0.39$ bit/s/Hz compared with AFDM with $\alpha=1.0$, confirming that the proposed SE advantage is preserved for the higher modulation order $M=16$.}

\begin{figure}[t]
	\centering		
	\subfigure[BLER performance.]{
		\label{BLER}
		\includegraphics[width=0.48\columnwidth]{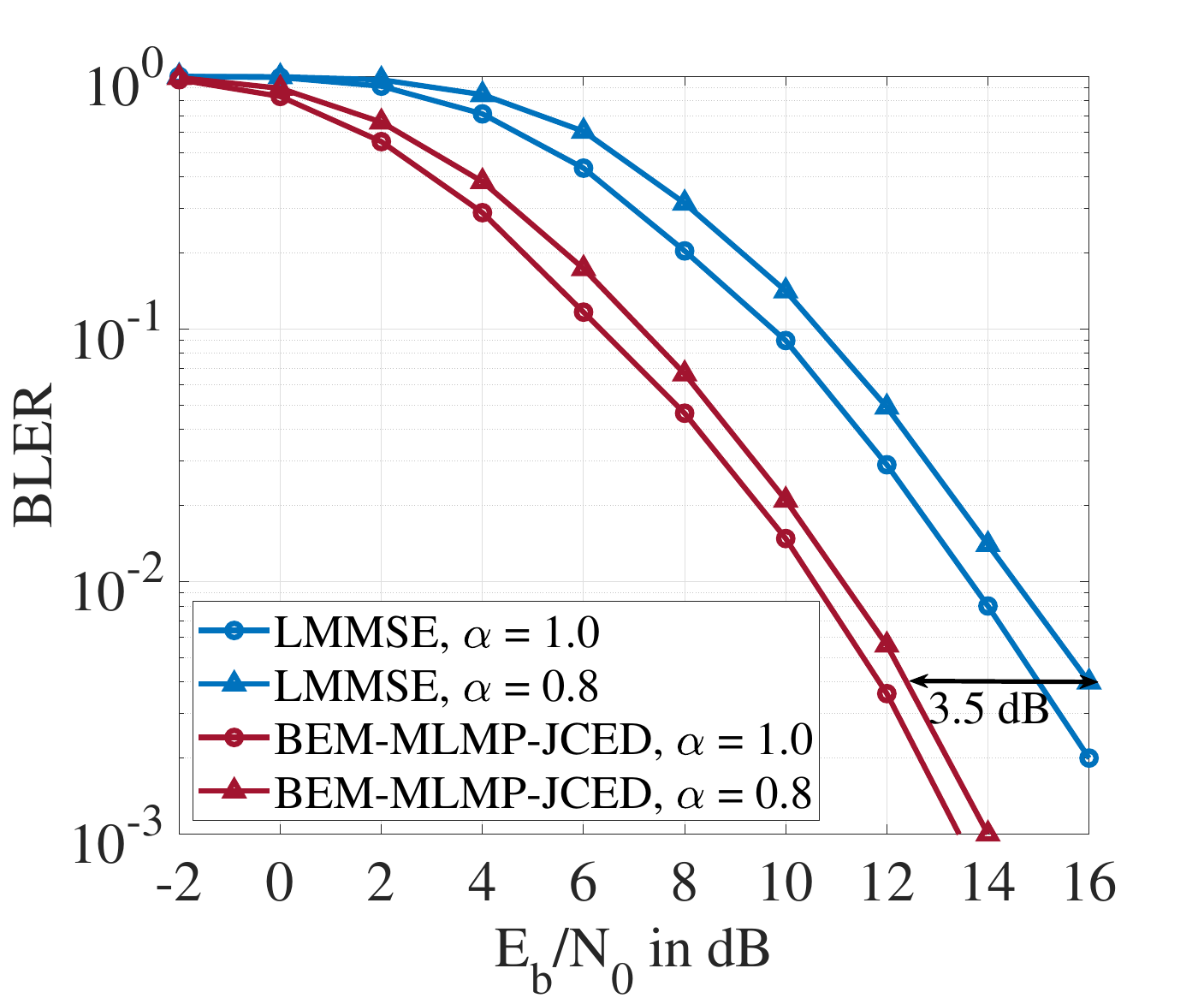}}	
	\subfigure[\textcolor{blue}{Practical effective SE.}]{
		\label{SEC}
		\includegraphics[width=0.48\columnwidth]{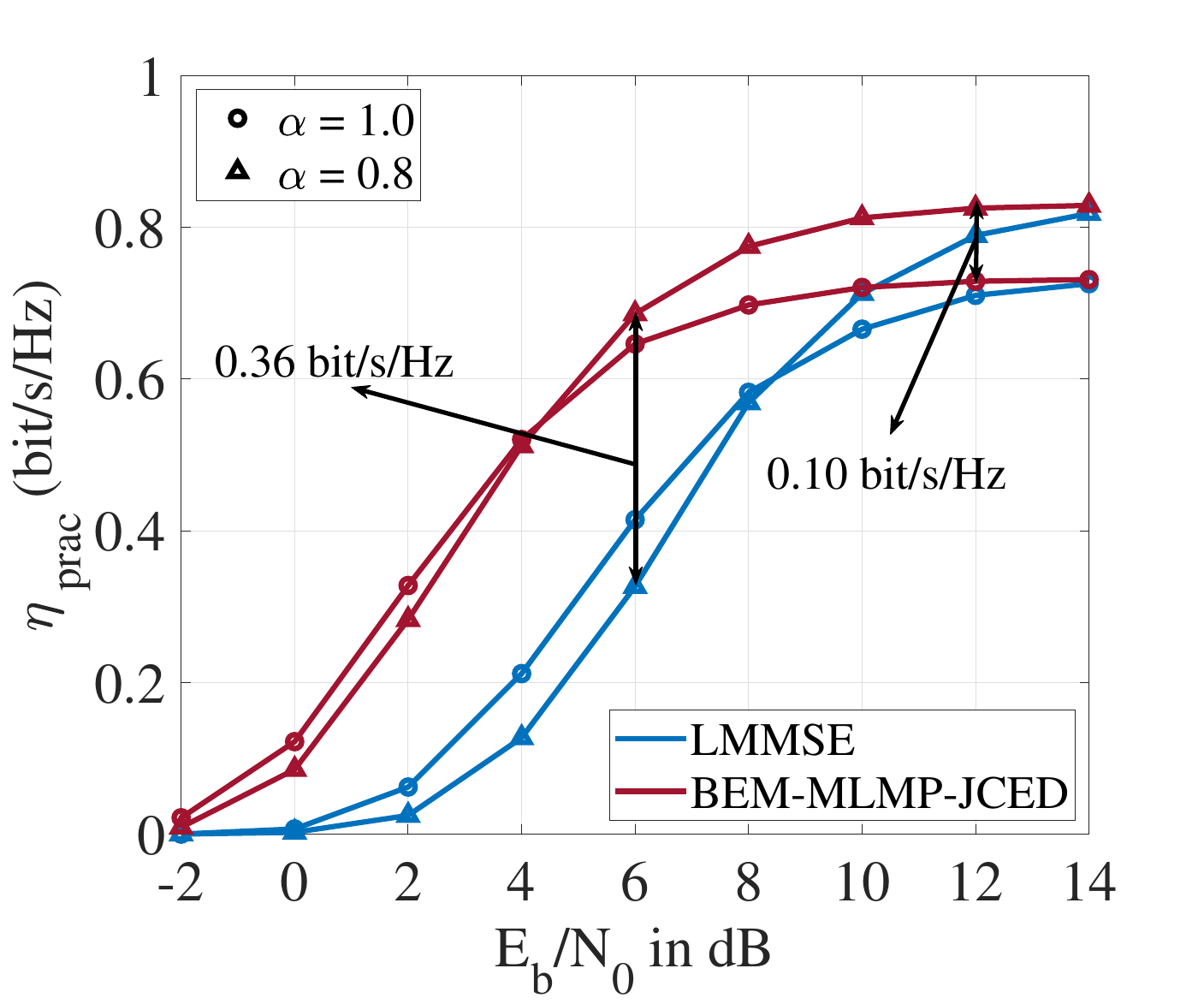}}		
	\caption{\textcolor{blue}{BLER and practical effective SE performance of rate-1/2 LDPC-coded AFDM-FTN system under different detection algorithms.}}
	\label{fig13}
    \vspace{-0.5cm}
\end{figure}

\textcolor{blue}{Fig. \ref{fig13} presents the block error rate (BLER) and practical effective SE performance of the rate-1/2 LDPC-coded AFDM-FTN system.  
Unlike the uncoded case, the practical effective SE of the coded AFDM-FTN system is evaluated by replacing the BER metric in \eqref{eq:eta_prac} with BLER.}  
\textcolor{blue}{From Fig. \ref{BLER}, it can be observed that the proposed BEM-MLMP-JCED algorithm still maintains a clear performance advantage in the coded systems.}
In particular, at a BLER of $4\times 10^{-3}$ with $\alpha = 0.8$, it achieves a gain of about $3.5$ dB over the LMMSE algorithm.
\textcolor{blue}{In terms of practical effective SE, Fig. \ref{SEC} shows that at $E_b/N_0 = 6$ dB with $\alpha = 0.8$, the proposed BEM-MLMP-JCED algorithm provides a gain of $0.36$ bit/s/Hz compared with LMMSE. Moreover, the proposed AFDM-FTN system achieves an additional $0.10$ bit/s/Hz SE gain over the conventional AFDM system at ${E_b/N_0} = 12$ dB.}
These results demonstrate that, in the coded system, our proposed schemes still exhibit clearly superior performance.

\begin{figure}[t]
	\centering
	\includegraphics[width=0.85\columnwidth]{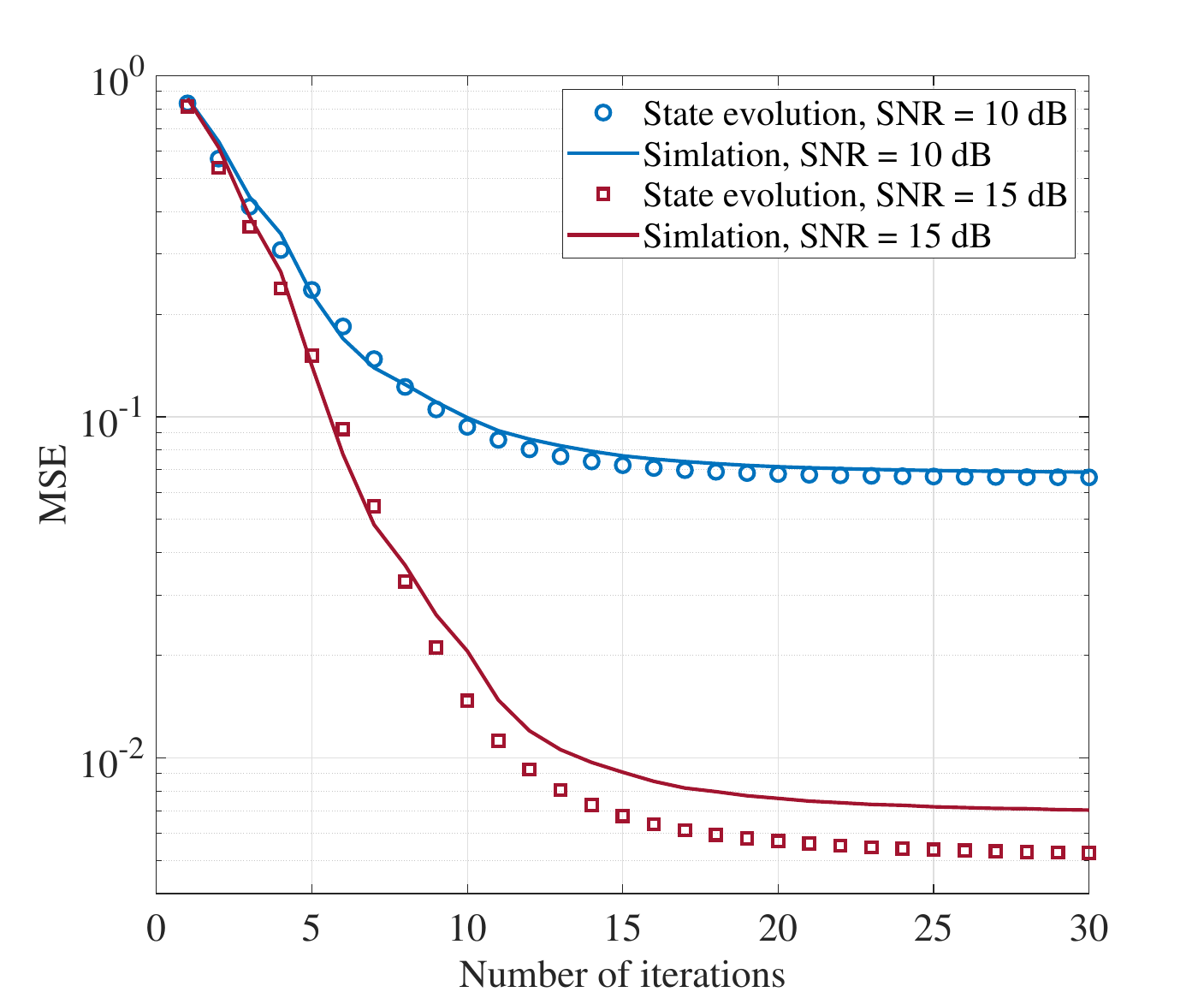}
	\caption{MSE of state evolution and simulation versus number of iterations for $\alpha = 0.8$.}
	\label{fig11}
    \vspace{-0.5cm}
\end{figure}

Fig. \ref{fig11} illustrates the MSE based state evolution of the proposed BEM-MLMP-JCED algorithm versus the number of iterations for different SNRs at $\alpha = 0.8$. As can be seen, the simulated MSE closely matches the MSE predicted by the state evolution analysis for all considered SNRs, which confirms the accuracy of the state evolution characterization and the effectiveness of the proposed BEM-MLMP-JCED.
\vspace{-0.5em}
\section{Conclusion}
\label{sec:7} 
This paper proposed an AFDM-FTN waveform to enhance SE in high-mobility communication environments. The input-output relationship of AFDM-FTN was first derived, showing that FTN induces coherent superposition across multipath components. To address the resulting channel estimation challenges, a BEM-assisted channel estimation method was   developed. In addition, by exploiting the sparsity of the TD channel matrix and the FTN coefficient matrix, we proposed a  MLMP detection algorithm, where the   message propagation   across the TD channel, FTN, and transform layers was discussed in detail. Building upon the BEM-assisted channel estimation and the MLMP detector, a BEM-MLMP-JCED scheme was further proposed to iteratively refine channel estimation with the aid of transmitted data. Theoretical analyses of the channel estimation lower bound, MSE performance via state evolution, and computational complexity were provided. Simulation results demonstrated that the proposed AFDM-FTN system with BEM-MLMP-JCED achieves lower BER than linear algorithms and OFDM-FTN counterparts with reduced complexity, while attaining comparable BER to conventional AFDM with enhanced SE.


\appendices
\vspace{-0.5em}
\section{}
\label{appe1}
With the least squares estimate of $\mathbf{c}$, the BEM model error can be expressed as a linear projection of $\mathbf{h}$, namely
\begin{equation}
  \scalebox{0.9}{$%
  \begin{aligned}
    \mathbf{e}_{\rm mod}
    = \Big(\mathbf{I}_{NL} - \mathbf{\tilde U}(\mathbf{\tilde U}^H\mathbf{\tilde U})^{-1}\mathbf{\tilde U}^H\Big)\mathbf{h}
    \triangleq \mathbf{\tilde U}_{\rm I}\mathbf{h}.
      \end{aligned}
  $}
  \label{eq:8081}
\end{equation}
In the Jakes' channel model, the channel response  vector $\mathbf{h}$ is zero-mean complex Gaussian. Since $\mathbf{e}_{\rm mod}$ is obtained from $\mathbf{h}$ through the deterministic linear transform $\mathbf{\tilde U}_{\rm I}$, it is also a zero-mean complex Gaussian vector.
For the $\ell$th path, the received error caused by model error is
\begin{equation}
  \scalebox{0.9}{$%
  \begin{aligned}
    \hat{\mathbf e}^{\,\ell}_{\rm mod}
    = \mathbf{A}\left(\mathbf e^{\ell}_{\rm mod}\odot\mathbf{v}_\ell\right)
    = \mathbf{A}\operatorname{diag}(\mathbf{v}_\ell)\mathbf e^{\ell}_{\rm mod},
  \end{aligned}
  $}
  \label{eq:808}
\end{equation}
where $\mathbf{v}_\ell = \mathbf{G}_\ell\mathbf{A}^H\mathbf{x}$, ${\mathbf { {e}}}^{\ell}_{\rm{mod}}$ represents the model error of the $\ell$th path. 
As a linear transform of $\mathbf{A}\operatorname{diag}(\mathbf{v}_\ell)$, ${\mathbf{\hat e}}^{\ell}_{\rm{mod}}$ is also a  complex Gaussian vector with zero-mean.
Stacking all paths as $\mathbf{\hat E}_{\rm{mod}}=\left[{\mathbf{\hat e}}^{0}_{\rm{mod}},{\mathbf{\hat e}}^{1}_{\rm{mod}},\cdots,{\mathbf{\hat e}}^{L-1}_{\rm{mod}} \right]^T$,
the covariance of $\mathbf{\hat E}_{\rm{mod}}$ can be expressed as
\begin{equation}
	\scalebox{0.88}{$%
		\begin{aligned}
			\mathbf{R}_{\rm{\hat E}} &=  \mathbb{E}\left[\mathbf{\hat E}_{\rm{mod}}^H\mathbf{\hat E}_{\rm{mod}} \right]
		  = \sum_{\ell=0}^{L-1} \bar P_\ell\|\mathbf v_\ell\|^2\ \mathbf A\left(\mathbf{ {U}}_{\rm{I}}\mathbf{R}_{\rm{h}}\mathbf{ {U}}_{\rm{I}}^H\right)\mathbf A^{H}.
		 \end{aligned}
		$}
	\label{eq:809}   
\end{equation}
Since $\mathbf{G}_\ell$ is a circulant matrix, it can be diagonalized by the DFT matrix as
$\mathbf{G}_\ell = \mathbf{F}^H\boldsymbol\Lambda_\ell\mathbf F$, where
$\boldsymbol\Lambda_\ell$ is a diagonal matrix with entries $\lambda_\ell[k]$, i.e., the eigenvalues of $\mathbf{G}_\ell$. 
Consequently, $\lVert \mathbf{v}_\ell \rVert^2$ can be computed as
\begin{equation}
	\scalebox{0.9}{$%
		\begin{aligned}
			\lVert \mathbf{v}_\ell \rVert^2
			=\lVert\boldsymbol\Lambda_\ell \mathbf F(\mathbf A^{H}\mathbf x)\rVert^2
			=\sum_{k=0}^{N-1} |\lambda_\ell[k]|^{2}|z_k|^{2}.
		\end{aligned}
		$}
	\label{eq:812}   
\end{equation}
Finally, we obtain
\begin{equation}
	\scalebox{0.9}{$%
		\begin{aligned}
			\sigma_{\rm{e}}^2 &=  \frac{1}{N}{\rm{Tr}}\left(\mathbf{R}_{\rm{\hat E}}\right) = \frac{1}{N}{\rm{Tr}}\left(\mathbf{ {U}}_{\rm{I}}\mathbf{R}_{\rm{h}}\mathbf{ {U}}_{\rm{I}}^H\right) \sum_{\ell=0}^{L-1}\bar P_\ell\sum_{k=0}^{N-1} |\lambda_\ell[k]|^{2}|z_k|^{2}.
		\end{aligned}
		$}
	\label{eq:813}   
\end{equation}

\vspace{-0.8em}

\section{}
\label{appe2}
We adopt a generic  back-projection to obtain pseudo observations as
\begin{equation}
	\scalebox{0.9}{$%
		\begin{aligned}
			 {\mathbf{x}}' = \hat{ \mathbf{x}}^{l} + \rho \mathbf{H}_{\rm{all}}^H\left(\mathbf{y}-\mathbf{H}_{\rm{all}}\hat{ \mathbf{x}}^{l}\right) 
		\end{aligned}.
		$}
	\label{eq:814}   
\end{equation}
Substituting into \eqref{eq72}, we get
\begin{equation}
	\scalebox{0.9}{$%
		\begin{aligned}
			 {\mathbf{x}}' - \mathbf{x} = \left( \mathbf{I}_N - \rho \mathbf{H}_{\rm{all}}^H \mathbf{H}_{\rm{all}}  \right) \mathbf{e} + \rho \mathbf{H}_{\rm{all}}^H\mathbf{w}.
		\end{aligned}
		$}
	\label{eq:815}   
\end{equation}
Assume that $\mathbf{e}$ is zero-mean, i.i.d., with covariance $\eta_x^{l}$, and is independent of $\mathbf{w}$, then the effective noise variance is
\begin{equation}
	\scalebox{0.9}{$%
\begin{aligned}
	\bar{\nu}^{l}_x(\rho)
	& = \frac{1}{N}\mathbb E\left[\|{\mathbf{x}}' - \mathbf{x}\|^2\right] \\
	&=\frac{1}{N}\mathbb E\left[\|(\mathbf {I}_N-\rho\mathbf{H}_{\rm{all}}^{H}\mathbf{H}_{\rm{all}})\mathbf e\|^2\right]
	+\frac{1}{N}\mathbb E\left[\|\rho \mathbf{H}_{\rm{all}}^{H}\mathbf w\|^2\right]\\
	&=\frac{\eta_x^{l}}{N}{\rm{Tr}}\left((\mathbf {I}_N-\rho\mathbf{H}_{\rm{all}}^{H}\mathbf{H}_{\rm{all}})^2\right)
	+\rho^2 \tau_{\rm{H_{all}}}\sigma^2.
\end{aligned}
		$}
\label{eq:816}   
\end{equation}
Based on orthogonalization calibration, we obtain $\bar{\nu}^{l}_x$ by taking $ \rho = \frac{1}{\tau_{\rm{H_{all}}}}$.
\vspace{-0.5em}
\bibliographystyle{IEEEtran}
\bibliography{reference.bib}

@article{luoDesignVariableModulation2025,
	title = {On the Design of Variable Modulation and Adaptive Modulation for Uplink Sparse Code Multiple Access},
	author = {Luo, Qu and Xiao, Pei and Chen, Gaojie and Zhu, Jing},
	year = 2025,
	month = apr,
	journal = {IEEE J. Sel. Areas Commun.},
	volume = {43},
	number = {4},
	pages = {1153--1167},
	issn = {1558-0008, 0733-8716}
}

@article{wuPerformanceAnalysisBEMBased2025,
  title = {Performance Analysis of {BEM}-Based Channel Estimation for {OTFS} With Hardware Impairments},
  author = {Wu, Haowei and Chen, Huanyu and Peng, Qihao and Luo, Qu and Ou, Jinglan},
  date = {2025-07},
  year={2025},
  month = jul,
  journal = {IEEE Commun. Lett.},
  volume = {29},
  number = {7},
  pages = {1719--1723},
  issn = {1089-7798, 1558-2558, 2373-7891},
  doi = {10.1109/LCOMM.2025.3572905}
}

@article{yi2025non,
  title={Non-Orthogonal Affine Frequency Division Multiplexing for Spectrally Efficient High-Mobility Communications},
  author={Yi, Qin and Liu, Zilong and Musavian, Leila and Sui, Zeping},
  url = {https://arxiv.org/abs/2508.09782},
  year={2025}
}

@ARTICLE{11185315,
  author={Sui, Zeping and Liu, Zilong and Musavian, Leila and Yang, Lie-Liang and Hanzo, Lajos},
  journal = {IEEE Trans. Wireless Commun.},
  title={Generalized Spatial Modulation Aided Affine Frequency Division Multiplexing}, 
  year={2025},
  month = Oct,
  volume={25},
  number={},
  pages={4658-4673},
  doi={10.1109/TWC.2025.3613062}
}

@article{zhuDesignPerformanceAnalysis2024,
  title = {Design and Performance Analysis of Index Modulation Empowered {AFDM} System},
  author = {Zhu, Jing and Luo, Qu and Chen, Gaojie and Xiao, Pei and Xiao, Lixia},
  date = {2024-03},
  year = {2024},
  month = mar,
  journal = {IEEE Wireless Commun. Lett.},
  volume = {13},
  number = {3},
  pages = {686--690},
  issn = {2162-2337, 2162-2345},
  doi = {10.1109/LWC.2023.3339704}
}

@article{bemaniAffineFrequencyDivision2023,
	title = {Affine Frequency Division Multiplexing for Next Generation Wireless Communications},
	author = {Bemani, Ali and Ksairi, Nassar and Kountouris, Marios},
	year = 2023,
	month = nov,
	journal = {IEEE Trans. Wireless Commun.},
	volume = {22},
	number = {11},
	pages = {8214--8229},
	issn = {1536-1276, 1558-2248},
	doi = {10.1109/TWC.2023.3260906}
}

@article{zhu2023design,
  title={Design and performance analysis of index modulation empowered {AFDM} system},
  author={Zhu, Jing and Luo, Qu and Chen, Gaojie and Xiao, Pei and Xiao, Lixia},
  journal={IEEE wireless commun.  lett.},
  volume={13},
  number={3},
  pages={686--690},
  year={2023},
month = Dec,
  publisher={IEEE}
}

@article{yinDiagonallyReconstructedChannel2024,
	title = {Diagonally Reconstructed Channel Estimation for {MIMO-AFDM} With Inter-{Doppler} Interference in Doubly Selective Channels},
	author = {Yin, Haoran and Wei, Xizhang and Tang, Yanqun and Yang, Kai},
	date = {2024-10},
	year = 2024,
	month = oct,
	journal = {IEEE Trans. Wireless Commun.},
	volume = {23},
	number = {10}
}

@article{11185309,
	author={Tao, Yiwei and Wen, Miaowen and Ge, Yao and Mao, Tianqi and Tang, Yanqun and Doosti-Aref, Abed},
	journal = {IEEE Trans. Wireless Commun.},
	title={Affine Frequency Division Multiple Access Based on {DAFT} Spreading for Next-Generation Wireless Networks}, 
   year={2025},
   month  = Oct,
  volume={25},
  number={},
  pages={4626-4641},
	doi={10.1109/TWC.2025.3612880}
	
}

@article{10858612,
	author={Ni, Yuanhan and Yuan, Peng and Huang, Qin and Liu, Fan and Wang, Zulin},
	journal = {IEEE Trans. Wireless Commun.},
	title={An Integrated Sensing and Communications System Based on Affine Frequency Division Multiplexing}, 
	year={2025},
	month = jan,
	volume={24},
	number={5},
	pages={3763-3779},
	doi={10.1109/TWC.2025.3532993}
}

@ARTICLE{11121670,
	author={Xia, Haowen and Zhang, Aihua and Tian, Di and Guo, Deshu},
	journal = {IEEE Commun. Lett.},
	title={An Embedded Single-Pilot-Aided Channel Estimation Scheme Based on Interference Position Indices for {AFDM-IM} in Delay-{Doppler} Channels}, 
	year={2025},
	month = aug,
	volume={29},
	number={10},
	pages={2381-2385},
	doi={10.1109/LCOMM.2025.3597205}}

@ARTICLE{11012133,
	author={Xia, Haowen and Zhang, Aihua and Guo, Deshu and Tian, Di and Wang, Shuangcheng},
	journal = {IEEE Wireless Commun. Lett.},
	title={A Single-Pilot-Aided Channel Estimation Scheme Based on Interference Position Indices for {AFDM} in Delay-{Doppler} Channels}, 
	year={2025},
	month = may,
	volume={14},
	number={8},
	pages={2466-2470},
	doi={10.1109/LWC.2025.3573104}
	}

@article{10439996,
	author={Bemani, Ali and Ksairi, Nassar and Kountouris, Marios},
	journal = {IEEE Wireless Commun. Lett.},
	title={Integrated Sensing and Communications With Affine Frequency Division Multiplexing}, 
	year={2024},
	month = feb,
	volume={13},
	number={5},
	pages={1255-1259},
	doi={10.1109/LWC.2024.3367178}
}

@article{11150613,
	author={Luo, Qu and Zhu, Jing and Liu, Zilong and Tang, Yanqun and Xiao, Pei and Chen, Gaojie and Shi, Jia},
	journal = {IEEE Trans. Wireless Commun.},
	title={Joint Sparse Graph for Enhanced {MIMO-AFDM} Receiver Design}, 
  year={2025},
month = Sep,
  volume={25},
  number={},
  pages={3272-3286},
	doi={10.1109/TWC.2025.3603012}
}

@article{liAffineFrequencyDivision2025a,
	title = {Affine Frequency Division Multiplexing Over Wideband Doubly-Dispersive Channels With Time-Scaling Effects},
	author = {Li, Xiangxiang and Wang, Haiyan and Ge, Yao and Shen, Xiaohong and Guan, Yong Liang and Wen, Miaowen and Yuen, Chau},
     year={2025},
month = Jul,
    volume={25},
    number={},
    pages={476-492},
	journal = {IEEE Trans. Wireless Commun.}
}

@article{luoAFDMSCMAPromisingWaveform2024,
	title = {{{AFDM-SCMA}}: {A} Promising Waveform for Massive Connectivity Over High Mobility Channels},
	author = {Luo, Qu and Xiao, Pei and Liu, Zilong and Wan, Ziwei and Thomos, Nikolaos and Gao, Zhen and He, Ziming},
	year={2024},
	month = oct,
	journal = {IEEE Trans. Wireless Commun.},
	volume = {23},
	number = {10},
	pages = {14421--14436},
	issn = {1536-1276, 1558-2248},
	doi = {10.1109/TWC.2024.3413980}
}

@article{ranasingheJointChannelData2025,
	title = {Joint Channel, Data, and Radar Parameter Estimation for {AFDM} Systems in Doubly-Dispersive Channels},
	author = {Ranasinghe, Kuranage Roche Rayan and  others},
	year={2025},
	month = feb,
	journal = {IEEE Trans. Wireless Commun.},
	volume = {24},
	number = {2},
	pages = {1602--1619},
	issn = {1536-1276, 1558-2248},
	doi = {10.1109/TWC.2024.3510935}
}

@article{sui2025multi,
  title={Multi-Functional Chirp Signalling for Next-Generation Multi-Carrier Wireless Networks: Communications, Sensing and {ISAC} Perspectives},
  author={Sui, Zeping and others},
  url={https://arxiv.org/abs/2508.06022},
  year={2025}
}

@article{wang2025afdm,
  title={{AFDM} Based Preamble Sequence Transmission for {6G} Mobile Satellite Communication Systems},
  author={Wang, Yanzhao and He, Yishan and Zhao, Lei and Jiang, Yuan},
  journal={IEEE Trans.    Wireless Commun.},
  year={2025},
  month = May,
  publisher={IEEE}
}

@misc{yin2025ofdm,
  title={From {OFDM} to {AFDM}: {Enabling} Adaptive Integrated Sensing and Communication in High-Mobility Scenarios},
  author={Yin, Haoran and others},
  url={https://arxiv.org/abs/2510.27192},
  year={2025}
}

@article{xiao2021overview,
  title={An overview of {OTFS} for Internet of Things: Concepts, benefits, and challenges},
  author={Xiao, Lixia and Li, Shuo and Qian, Ying and Chen, Da and Jiang, Tao},
  journal={IEEE Internet   Things J. },
  volume={9},
  number={10},
  pages={7596--7618},
  year={2021},
  month = Dec,
  publisher={IEEE}
}

@article{liChirpParameterSelection2025,
title = {Chirp Parameter Selection for Affine Frequency Division Multiplexing With {MMSE} Equalization},
author = {Li, Zunqi and Zhang, Chuanbin and Song, Ge and Fang, Xiaojie and Sha, Xuejun and Slock, Dirk T. M.},
date = {2025-07},
year={2025},
month = JUL,
journal = {IEEE Trans. Commun.},
volume = {73},
number = {7},
pages = {5079--5093},
issn = {0090-6778, 1558-0857},
doi = {10.1109/TCOMM.2024.3519521}
}

@article{deng2025unifying,
  title={A unifying view of {OTFS} and its many variants},
  author={Deng, Qinwen and Ge, Yao and Ding, Zhi},
  journal={IEEE Commun.  Surveys \& Tut.},
  year={2025},
  month = Feb,
  publisher={IEEE}
}

@article{lin2022orthogonal,
  title={Orthogonal delay-Doppler division multiplexing modulation},
  author={Lin, Hai and Yuan, Jinhong},
  journal={IEEE Trans.    Wireless Commun.},
  volume={21},
  number={12},
  pages={11024--11037},
  year={2022},
  month = Jul,
  publisher={IEEE}
}

@article{ouyang2016orthogonal,
  title={Orthogonal chirp division multiplexing},
  author={Ouyang, Xing and Zhao, Jian},
  journal={IEEE Trans.  on Commun.},
  volume={64},
  number={9},
  pages={3946--3957},
  year={2016},
month = Jul,
  publisher={IEEE}
}

@article{prljaReducedComplexityReceiversStrongly2012,
	title = {Reduced-Complexity Receivers for Strongly Narrowband Intersymbol Interference Introduced by Faster-than-{Nyquist} Signaling},
	author = {Prlja, Adnan and Anderson, John B.},
	date = {2012-09},
	year = {2012},
	month = sep,
	journal = {IEEE Trans. Commun.},
	volume = {60},
	number = {9},
	pages = {2591--2601},
	issn = {0090-6778},
	doi = {10.1109/TCOMM.2012.070912.110296}
}

@article{liverisExploitingFasterthannyquistSignaling2003,
	title = {Exploiting Faster-than-{Nyquist} Signaling},
	author = {Liveris, A.D. and Georghiades, C.N.},
	year = {2003},
	month = sep,
	journal = {IEEE Trans. Commun.},
	volume = {51},
	number = {9},
	pages = {1502--1511},
	issn = {0090-6778},
	doi = {10.1109/TCOMM.2003.816943}
}

@article{yuanIterativeReceiverDesign2020,
	title = {Iterative Receiver Design for {FTN} Signaling Aided Sparse Code Multiple Access},
	author = {Yuan, Weijie and Wu, Nan and Zhang, Andrew and Huang, Xiaojing and Li, Yonghui and Hanzo, Lajos},
	date = {2020-02},
	year = {2020},
	month = feb,		
	journal = {IEEE Trans. Wireless Commun.},
	volume = {19},
	number = {2},
	pages = {915--928},
	issn = {1536-1276, 1558-2248},
	doi = {10.1109/TWC.2019.2950000}
}

@article{tongAdaptiveFTNSignaling2025,
	title = {Adaptive {FTN} Signaling Over Rapidly-Fading Channels},
	author = {Tong, Mingfei and Huang, Xiaojing and Andrew Zhang, J. and Hanzo, Lajos},
	date = {2025-09},
	year = {2025},
	month = sep,	
	journal = {IEEE Trans. Commun.},
	volume = {73},
	number = {9},
	pages = {7166--7178},
	issn = {0090-6778, 1558-0857},
	doi = {10.1109/TCOMM.2025.3545655}
}

@article{hongPrecodedFasterThanNyquistSignaling2025,
	title = {Precoded Faster-Than-{Nyquist} Signaling Using Optimal Power Allocation for {OTFS}},
	author = {Hong, Zekun and Sugiura, Shinya and Xu, Chao and Hanzo, Lajos},
	date = {2025-01},
	year = {2025},
	month = jan,	
	journal = {IEEE Wireless Commun. Lett.},
	volume = {14},
	number = {1},
	pages = {173--177},
	issn = {2162-2337, 2162-2345},
	doi = {10.1109/LWC.2024.3491777}
}

@article{liuNearOptimalBEMOTFS2022,
	title = {Near-Optimal {BEM OTFS} Receiver With Low Pilot Overhead for High-Mobility Communications},
	author = {Liu, Yujie and Guan, Yong Liang and G., David Gonzalez},
	date = {2022-05},
	year = {2022},
	month = May,		
	journal = {IEEE Trans. Commun.},
	volume = {70},
	number = {5},
	pages = {3392--3406},
	issn = {0090-6778, 1558-0857},
	doi = {10.1109/TCOMM.2022.3162257}
}

@article{9369968,
	author={Liu, Zilong and Yang, Lie-Liang},
	journal = {IEEE Trans. Wireless Commun.},
	title={Sparse or Dense: {A} Comparative Study of Code-Domain {NOMA} Systems}, 
	year={2021},
	month = mar,
	volume={20},
	number={8},
	pages={4768-4780},
	doi={10.1109/TWC.2021.3062235}}

@article{pakroohAnalysisFisherInformation2015,
  title = {Analysis of {Fisher} Information and the {Cramér–Rao} Bound for Nonlinear Parameter Estimation After Random Compression},
  author = {Pakrooh, Pooria and Pezeshki, Ali and Scharf, Louis L. and Cochran, Douglas and Howard, Stephen D.},
  date = {2015-12},
  year  = {2015},
  month = Dec,
  journal = {IEEE Trans. Signal Process.},
  volume = {63},
  number = {23},
  pages = {6423--6428},
  issn = {1053-587X, 1941-0476},
  doi = {10.1109/TSP.2015.2464183}
}

@article{liTimeDomainVsFrequencyDomain2020,
  title = {Time-Domain vs. Frequency-Domain Equalization for {FTN} Signaling},
  author = {Li, Shuangyang and Yuan, Weijie and Yuan, Jinhong and Bai, Baoming and Wing Kwan Ng, Derrick and Hanzo, Lajos},
  year = 2020,
  month = aug,
  journal = {IEEE Trans. Veh. Technol.},
  volume = {69},
  number = {8},
  pages = {9174--9179},
  issn = {0018-9545, 1939-9359},
  doi = {10.1109/TVT.2020.3000074}
}

@article{liCodeBasedChannelShortening2020,
  title = {Code-Based Channel Shortening for Faster-Than-{Nyquist} Signaling: {Reduced-Complexity Detection} and Code Design},
  shorttitle = {Code-{{Based Channel Shortening}} for {{Faster-Than-Nyquist Signaling}}},
  author = {Li, Shuangyang and Yuan, Jinhong and Bai, Baoming and Benvenuto, Nevio},
  date = {2020-07},
  year = {2020},
  month = Jul,
  journal = {IEEE Trans. Commun.},
  volume = {68},
  number = {7},
  pages = {3996--4011},
  issn = {0090-6778, 1558-0857},
  doi = {10.1109/TCOMM.2020.2988922}
}

@article{pengResourceAllocationUplink2023,
  title = {Resource Allocation for Uplink Cell-Free Massive {MIMO} Enabled {URLLC} in a Smart Factory},
  author = {Peng, Qihao and Ren, Hong and Pan, Cunhua and Liu, Nan and Elkashlan, Maged},
  year = {2022},
  month = Nov,
  journal = {IEEE Trans. Commun.},
  volume = {71},
  number = {1},
  pages = {553--568},
  issn = {0090-6778, 1558-0857},
  doi = {10.1109/TCOMM.2022.3224502}
}

@article{rouAffineFrequencyDivision2026,
  title = {Affine Frequency Division Multiplexing ({{AFDM}}) for {{6G}}: {{Properties}}, Features, and Challenges},
  author = {Rou, Hyeon Seok and Ranasinghe, Kuranage Roche Rayan and Savaux, Vincent and Abreu, Giuseppe Thadeu Freitas De and David González, G. and Masouros, Christos},
	note  =  {{Early Access}, {DOI}: {10.1109/MCOMSTD.2025.3643183}, 2026},
  journal = {IEEE Comm. Stand. Mag.},
}

\end{document}